\documentclass[a4paper,fleqn,usenatbib]{mnras}
\usepackage{float}
\usepackage[T1]{fontenc}
\usepackage{ae,aecompl}
\usepackage{graphicx}
\usepackage{amsmath}
\usepackage{amssymb}
\usepackage{supertabular}
\usepackage{siunitx}
\usepackage{caption}

\def\beq{\begin{eqnarray}}
\def\eeq{\end{eqnarray}}

\title[Global solutions of MCNDAFs]{Relativistic global solutions of neutrino-dominated accretion flows with magnetic coupling}

\author[She, Liu, \& Xue]{Jiao-Zhen She, Tong Liu\thanks{tongliu@xmu.edu.cn}, Li Xue\thanks{lixue@xmu.edu.cn}\\
Department of Astronomy, Xiamen University, Xiamen, Fujian 361005, China}

\date{Accepted XXX. Received YYY; in original form ZZZ}

\pubyear{2022}

\begin{document}

\maketitle

\begin{abstract}
A Kerr black hole (BH) surrounded by a neutrino-dominated accretion flow (NDAF) is one of plausible candidates of the central engine in gamma-ray bursts. The accretion material might inherit and restructure strong magnetic fields from the compact object mergers or massive collapsars. The magnetic coupling (MC) process between a rapid rotating BH and an accretion disc is one of possible magnetic configurations that transfers the energy and angular momentum from the BH to the disc. In this paper, we investigate one-dimensional global solutions of NDAFs with MC (MCNDAFs), taking into account general relativistic effects, detailed neutrino physics, different MC geometries, and reasonable nucleosynthesis processes. Six cases with different accretion rates and power-law indices of magnetic fields are presented and compared with NDAFs without MC. Our results indict that the MC process can prominently impact the structure, thermal properties, and microphysics of MCNDAFs, increase luminosities of neutrinos and their annihilations, result in the changing of radial distributions of nucleons, and push the region of heavy nuclei synthesis to a larger radius than counterparts in NDAFs.
\end{abstract}

\begin{keywords}
accretion, accretion discs - black hole physics - gamma-ray burst: general - magnetic fields - neutrinos - nuclear reactions, nucleosynthesis, abundances
\end{keywords}

\section{Introduction} \label{sec:intro}

Extraction of energy and angular momentum from a spinning black hole (BH) or an accretion disc through magnetic braking is a long-standing topic. There are several possible types of magnetic field geometries associated with this issue \citep[e.g.][]{Blandford2002,Hirose2004,McKinney2005}. For example, small-scale tangled fields embedded in a disc is related to the turbulent viscosity of accretion flow through the Balbus-Hawley instability \citep{Balbus1991}, large-scale loop fields connecting different regions of a disc will result in the long-distance interactions between these regions \citep[e.g.][]{Blandford2002,Hirose2004,McKinney2005}, and large-scale open fields emerging from a disc will lead to winds through Blandford-Payne (BP) process \citep{Payne1982}.

However, none of above geometries is directly related to BHs. \citet{Blandford1977} proposed the first field geometry directly linked to BHs, which consists of large-scale open fields emerging from the BH and ending at the remote astrophysical load, and by which the energy and angular momentum can be extracted from the BH and transported to distance via the Poynting flux. This mechanism is named as the Blandford-Znajek (BZ) mechanism. BZ and BP mechanisms had been widely considered as origins of jets or outflows (winds) from BH binaries, active galactic nuclei, and gamma-ray bursts \citep[GRBs, e.g.][]{Blandford1977,Macdonald1982,Rees1984,Cao1994,Cao2013,Lee2000a,Lee2000b,Piran2004,Remillard2006,Barkov2008,Cao2011}. In order to investigate interactions between a BH and a disc, the field lines connecting the BH and disc through the plunging region were considered in a series of works \citep[e.g.][]{Krolik1999,Gammie1999,Agol2000,Hawley2000,Hawley2001}. Nevertheless, large-scale fields through the disc corona rather than the unstable plunging region will lead to more intense interactions between the BH and disc, which is generally called magnetic coupling (MC) process \citep[e.g.][]{Lip2000,Li2002}. This process has been generally adopted in researches of different classic accretion flows around BHs, such as the thin disc \citep[e.g.][]{Li2002,Wang2002,Wang2003a,Wang2003b,U2005,kluzniak2007}, the slim disc \citep[e.g.][]{Lee1999}, and the advection-dominated accretion flow \citep[e.g.][]{Ma2007}. A common conclusion amount these works is that the MC process will largely enhance the radiation luminosity by the additional energy extracted from BHs, meanwhile the disc structure will be significantly changed by the additional angular momenta from BHs.

As one of candidates of GRB central engines, the neutrino-dominated accretion flow (NDAF) has been used to explain many characteristics of GRBs in past two decades, such as GRB jet luminosities powered by the annihilations of neutrinos escaping from the discs \citep[e.g.][]{Popham1999,Fryer1999,Lee2005,Gu2006,Liu2007,Chen2007,Zalamea2011,Lei2013,Xue2013,Xie2016,Song2020}, variabilities in GRB prompt emissions driven by the instabilities of the discs \citep[e.g.][]{Reynoso2006,Janiuk2007,Kawanaka2007,Lei2009,Lei2017,Liu2010,Caballero2011,Kawanaka2012,Kawanaka2013,Hou2014,Kimura2015,Lin2016,Kawanaka2019,Shahamat2021}, extended emissions following short-duration GRBs induced by the redistribution of the angular momenta of the discs \citep[e.g.][]{Liu2012a,Cao2014}, X-ray flares originated from the instability or magnetic barriers in the discs \citep[e.g.][]{Perna2006,Proga2006,Lazzati2008,Liu2008,Lee2009,Luo2013,Mu2016,Mu2018,Yi2022}, brightly optical bumps of the core-collapse supernovae (CCSNe) associated with GRBs supplied by the disc outflows \citep[e.g.][]{Surman2006,Surman2011,Song2019,Liu2021}, long-lived CCSNe with at least two peaks produced by the disc outflow feedback \citep{Liu2019}, luminous kilonovae driven by outflows (or the gamma-ray radiation induced by the vertical advection) in binary compact object mergers \citep[e.g.][]{Yi2017,Yi2018,Song2018,Qi2022}, and possibly detectable MeV neutrinos and gravitational waves \citep[GWs, e.g.][]{Romero2010,Sun2012,Pan2012,Caballero2016,Xie2016,Liu2017a,Wei2019,Wei2020,Wei2021}. For the recent reviews, see \citet{Liu2017b} and \citet{Zhang2018}.

The MC process is one of practicable mechanisms in accretion discs including NDAFs. \citet{Lei2009} investigated the properties of magnetic coupling NDAFs (MCNDAFs) and discussed their thermal-viscous instabilities to interpret variabilities on GRB light curves. \citet{Luo2013} studied MCNDAFs to investigate the observed late-time X-ray flares of GRBs. Moreover, \citet{Song2020} discussed MeV neutrinos and corresponding GWs from a magnetized NDAF including the MC process. The results all confirmed that the MC process can significantly enhance the neutrino and neutrino annihilation luminosities. However, these works are all simplified local algebraic solutions, and the detailed dynamics and neutrino physics are excluded. Hence relativistic global solutions of MCNDAFs should be visited.

In this paper, the relativistic global solutions of the axisymmetric MCNDAF in Kerr metric are calculated. In order to highlight the MC process, we ignore any possible BZ and BP processes also related to large-scale magnetic fields. In section \ref{Section_Model}, we establish the basic model by introducing the fundamental hydrodynamic and thermodynamic equations with the detailed neutrino physics and nuclear statistical equilibrium \citep[NSE, e.g.][]{Seitenzahl2008,Liu2013,Xue2013}. In section \ref{sec_num_res}, the numerical method and results are exhibited. Conclusions and relevant discussions are made in section \ref{sec_con_dis}.

\section{Model}\label{Section_Model}

Catastrophic events from high-angular-momentum compact sources such as massive collapsars or BH-neutron star mergers are expected to form a torus. However, a thick torus is generally unstable \citep{Papaloizou1984} and it will undergo the self-gravity instability. If the torus with non-axisymmetric distortions reaches an appreciable mass fraction of the central BH, the instability may remove the non-axisymmetry from the accretion flow \citep{Woodward1994}. This makes the situation possible for a stable axisymmetric inner disc plus a instable non-axisymmetric outer thick torus. In this paper, we only focus on the possible MC process of axisymmetric inner disc, which is widely considered as one of plausible GRB central engine candidates \citep[for a review see][]{Liu2017b}. The research on interactions between an outer torus and an inner disc has gone beyond the scope of this paper, but it is an interesting topic worthy of being involved in future research. We take account of general relativity effects, neutrino physics, magnetic field geometries, and nucleosynthesis processes in NDAF model based on our previous works \citep{Xue2013,Liu2017b}. In order to make this paper reader friendly, we revisit most of these physics in details here.

\subsection{Relativistic hydrodynamics}

The relativistic hydrodynamics in our model is refered to the ones in the ADAF \citep{Abramowicz1996}, NDAF \citep{Popham1999}, and slim disc \citep{Sadowski2009} models, which are all the one-dimensional global solutions in the Kerr metric. To simplify, the hydrodynamics in our model is described in units of $G = c = M_{\rm BH} = 1$ ($M_{\rm BH}$ is the BH mass), but the neutrino physics and thermodynamics is described in cgs units. The five basic hydrodynamics equations are as follows.

First, the mass conservation equation is
\beq \label{continuity}
\dot{M}=-4\pi\rho H \Delta^{1/2}\frac{V_r}{\sqrt{1-{V_r}^2}},
\eeq
where $\dot{M}$, $\rho$, $H$, and $V_r$ are the rest-mass accretion rate, the rest-mass density, the half-thickness of disc, and the radial velocity measured in the corotating frame, respectively. Note that $\Delta=r^2-2r+a^2$ is a function of the Boyer-Lindquist radial coordinate $r$, and $a$ is the total specific angular momentum of the BH.

Second, the gas energy equation is written as
\beq \label{energy}
-\frac{\dot{M}}{2\pi r^2} \left(\frac{u}{\rho}\frac{d\ln u}{d\ln r}- \frac{p}{\rho}\frac{d\ln\rho}{d\ln r}\right)=-\frac{2\alpha p H A \gamma^2}{r^3}\frac{d\Omega}{dr}\nonumber\\-Q^-+Q_{\rm MC},
\eeq
where $u$, $p$, and $\alpha$ is the specific internal energy, pressure, and viscosity parameter on the disc \citep[see e.g.][]{Kato2008}, respectively. The variable-$\alpha$ effect could incur the instability of the disc \citep[e.g.][]{Simon2009,Potter2014,Kawanaka2019}. Moreover, $A\equiv r^4+r^2 a^2+2 r a^2$, $\gamma$ is the Lorentz factor, and $\Omega\equiv u^{\phi}/u^{t}$ is the angular velocity with respect to the stationary observer. It is noted that the l.h.s of this equation corresponds to the advection cooling rate $Q_{\rm adv}$ and the first term on the r.h.s is the viscous heating rate $Q_{\rm vis}$. The terms of $Q^-$ and $Q_{\rm MC}$ are respectively the total cooling rate and the energy injection through MC process, which are both described in Section \ref{sec_thermodynamics}. Thus this equation represents the balance between the coolings and heatings of the gas internal energy.

Third, the equation of angular momentum conservation is
\beq \label{angular-momentum}
\dot{M}(\mathcal{L}-\mathcal{L}_{\rm{in}})+T_{\rm MC}=\frac{4\pi p H A^{1/2}\Delta^{1/2}\gamma}{r},
\eeq
where $\mathcal{L}\equiv u_\phi$ and $\mathcal{L}_{\rm in}$ are the specific angular momentum of the accreting gas and specific angular momentum at the inner boundary of the disc, respectively. $T_{\rm MC}$ is the MC torque, which is described in Section \ref{sec_thermodynamics}.

Fourth, the equation of radial momentum conservation is
\beq \label{radial-momentum}
\frac{V_r}{1-{V_r}^2}\frac{d{V_r}}{dr}=\frac{\mathcal{A}}{r}-(1-{V_r}^2)\frac{1}{\lambda\rho}\frac{dp}{dr},
\eeq
where
\beq \mathcal{A}\equiv-\frac{A}{r^3\Delta\Omega^+_{\rm K}\Omega^-_{\rm K}}\frac{(\Omega-\Omega^+_{\rm K})(\Omega-\Omega^-_{\rm K})}{1-\tilde{\Omega}^2\tilde{R}^2}.
\eeq
The effects of gravity and rotation are combined in $\mathcal{A}$, where $\lambda\equiv(\rho+p+u)/\rho$ is the relativistic enthalpy, $\tilde{\Omega}\equiv(\Omega-2ar/A)$ is the angular velocity relative to the local inertial observer, $\Omega^{\pm}_{\rm K}\equiv\pm(r^{3/2}\pm a)^{-1}$ are the angular frequencies of the corotating and counterrotating Keplerian orbits, and $\tilde{R}\equiv A/(r^2 \Delta^{1/2})$ is the radius of gyration.

Fifth, the equation of vertical mechanical equilibrium is \citep[e.g.][]{Abramowicz1997,Xue2013,Liu2017b}
\beq\label{vertical-eq}
\frac{p}{\lambda\rho H^2}=\frac{\mathcal{L}^2-a^2(\epsilon^2-1)}{r^4},
\eeq
where $\epsilon\equiv u_t$ is the energy at infinity along geodesics. In our calculation, the formulae of $\epsilon$ and $\gamma$ are respectively defined as
\beq
\epsilon=-\gamma\frac{r\Delta^{1/2}}{A^{1/2}}-\frac{2ar}{A}\mathcal{L}
\eeq
and
\beq
\gamma=\sqrt{\frac{1}{1-{V_r}^2}+\frac{\mathcal{L}^2r^2}{A}}.
\eeq

\subsection{Neutrino physics}

The cooling mechanism is the main difference between NDAFs and classic accretion discs. Neutrino radiation becomes dominant in NDAFs as well as MCNDAFs. The precise neutrino physics is considered in our calculation and the details are described in following.

\subsubsection{Neutrino optical depth}

The total optical depth of neutrinos is
\beq
\tau_{\nu_i}=\tau_{s, \nu_i} + \tau_{a,\nu_i},
\eeq
which consists of two parts from scattering and absorption, denoted by subscript ``$s$'' and ``$a$'', respectively. The symbol $\nu_i$ represents the three different species of neutrinos $\nu_{\rm e}$, $\nu_{\rm \mu}$, and $\nu_{\rm \tau}$, as well as for the antineutrinos.

The optical depth for neutrinos through scattering off electrons and nucleons $\tau_{s, \nu_i}$ is given by
\beq
\tau_{s,\nu_i} \approx H \displaystyle(\sigma_{{\rm e^-}, \nu_i} n_{\rm e^-} + \sigma_{{\rm e^+}, \nu_i} n_{\rm e^+}+ {\sum_{j}} \sigma_{j,\nu_i} n_j),
\eeq
where $\sigma_{\rm e^{\mp}, \nu_i}$ and $\sigma_{j, \nu_i}$ are the cross sections of electron (positron) and nucleons \citep[e.g.][]{Kohri2005,Chen2007,Xue2013,Liu2017b}. $n_{\rm e^{\mp}}$ is the number density of electron (positron); based on the Fermi-Dirac integration \citep[see, e.g.][]{Kohri2005,Kawanaka2007,Liu2007,Liu2017b}, the number densities of electrons and positron $n_{\rm e^-}$ and $n_{\rm e^+}$ can be obtained, i.e.,
\beq
n_{\rm e^{\mp}}= \frac{1}{\pi^2 \hbar^3} \int_0^\infty d p \frac{p^2}{{\rm e}^{({\sqrt{p^2 c^2+{m_{\rm e}}^2 c^4} \mp {\mu_{\rm e}})/k_{\rm B} T}}+1},
\eeq
where $\mu_{\rm e}=\eta_{\rm e} k_{\rm B} T$ is the chemical potential of electrons, $\eta_{\rm e}$ is the electron degeneracy, and $T$ is the temperature of the disc. $n_j$ ($j =$ 1, 2, ...) are the number densities of nucleons ($j=1$ for free proton, $j=2$ for free neutron, and $j\ge3$ for other heavy nucleons), which are obtained by the NSE as shown in Section \ref{sec_nucleosynthesis}. From a given $n_j$, the mass fraction of the $j$th nucleons can be derived.

Moreover, the absorption depth for neutrinos $\tau_{a, \nu_i}$ is defined as
\beq
\tau_{a, \nu_i}= \frac{q_{\nu_i} H}{4 (7/8) \sigma T^4},
\eeq
where $q_{\nu_i}$ is the total neutrino cooling rate per unit volume, which consists of four parts, i.e., the Urca process, the electron-positron pair annihilation rate, the nucleon-nucleon bremsstrahlung rate, and the plasmon decay rate,
\beq
q_{\nu_i}=q_{\rm Urca}+q_{{\rm e^{-}+e^{+}} \rightarrow \nu_i+ \overline {\nu}_i}+q_{{\rm n + n \rightarrow  n + n +} \nu_i+\overline{\nu}_i}\nonumber\\+q_{\tilde{\gamma} \rightarrow \nu_i+\overline \nu_i},
\eeq
and more detailed physical descriptions see \citet{Xue2013} and \citet{Liu2017b}.

\begin{figure*}
\centering
\includegraphics[width=0.45\textwidth]{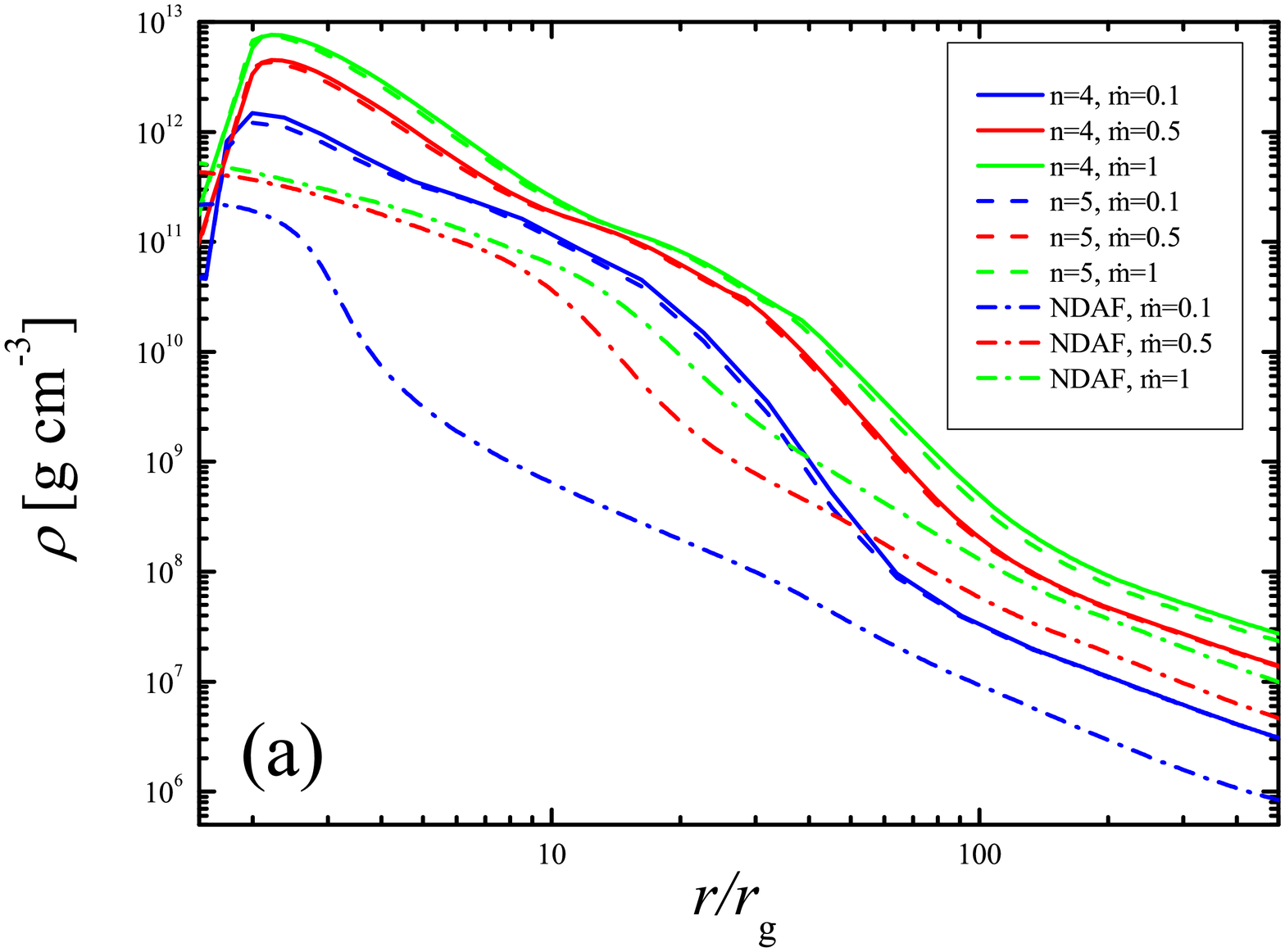}
\includegraphics[width=0.45\textwidth]{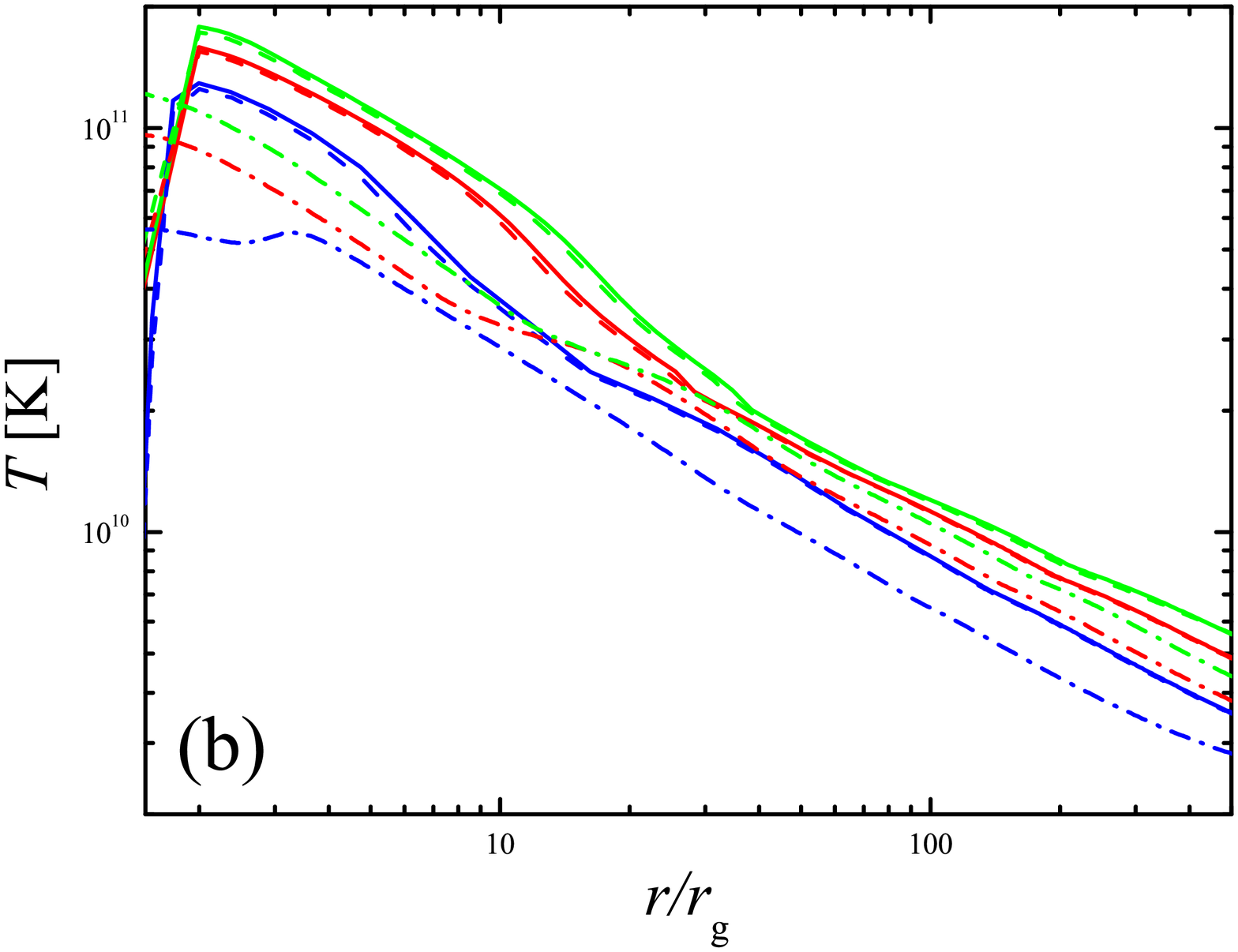}
\includegraphics[width=0.45\textwidth]{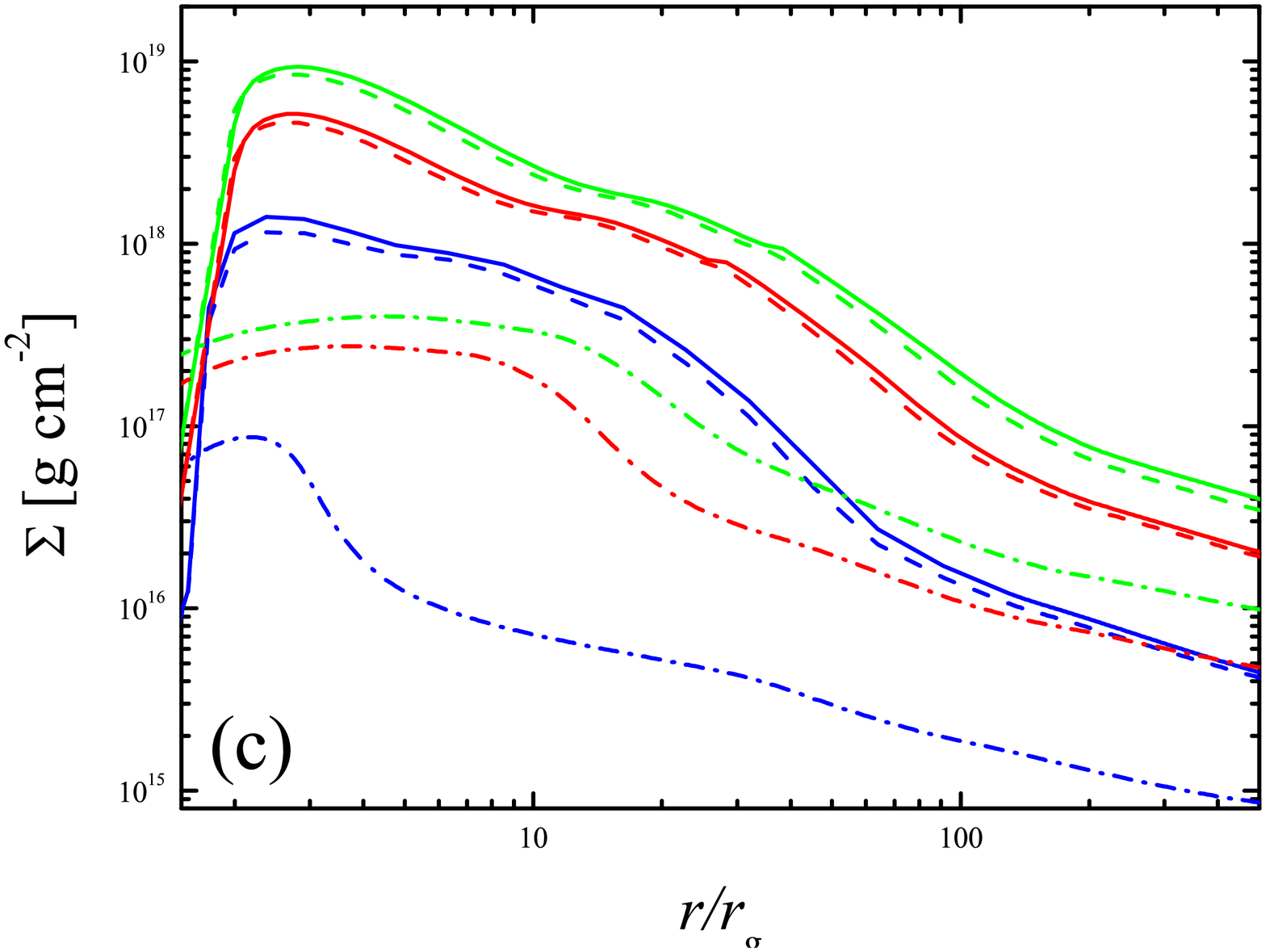}
\includegraphics[width=0.45\textwidth]{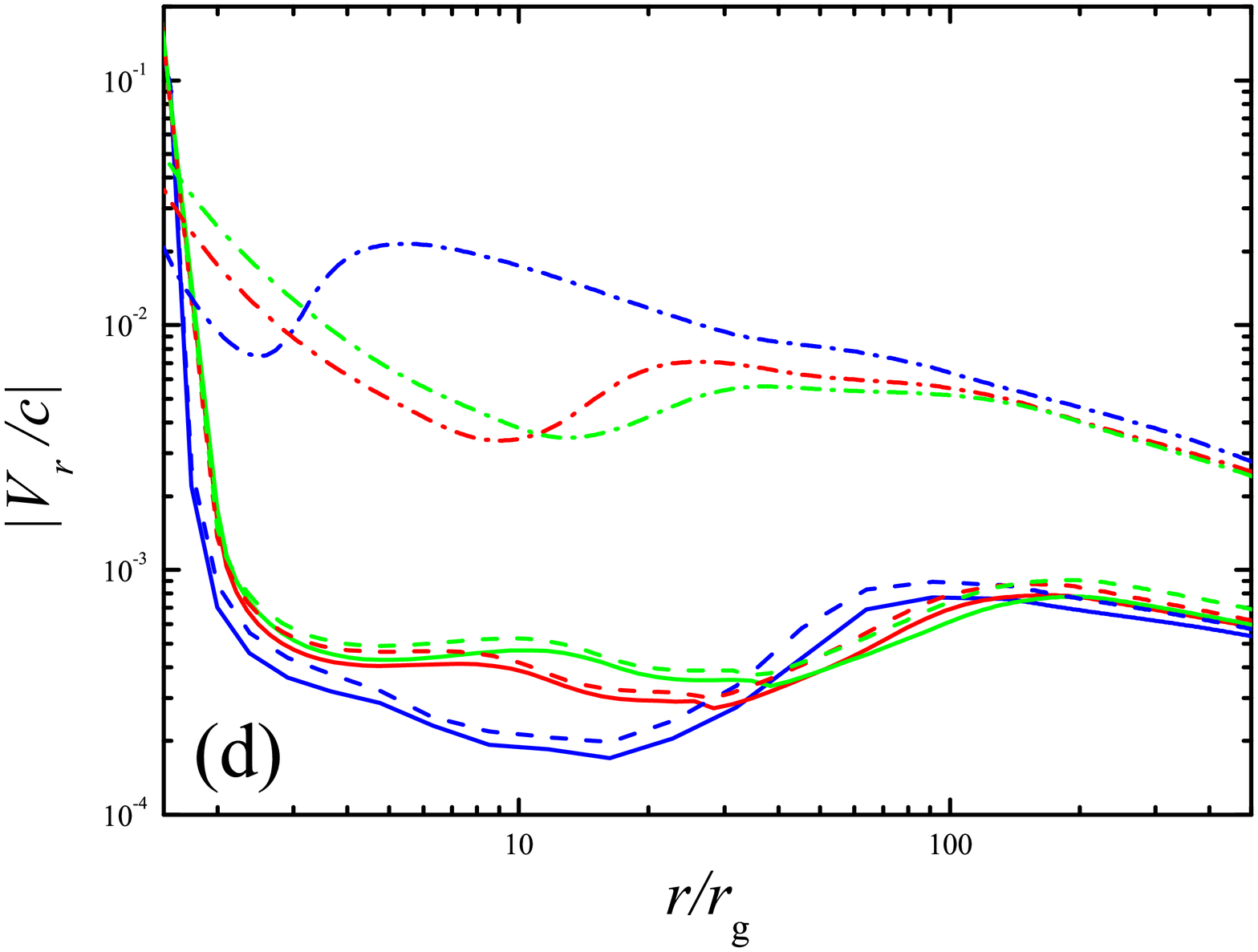}
\includegraphics[width=0.45\textwidth]{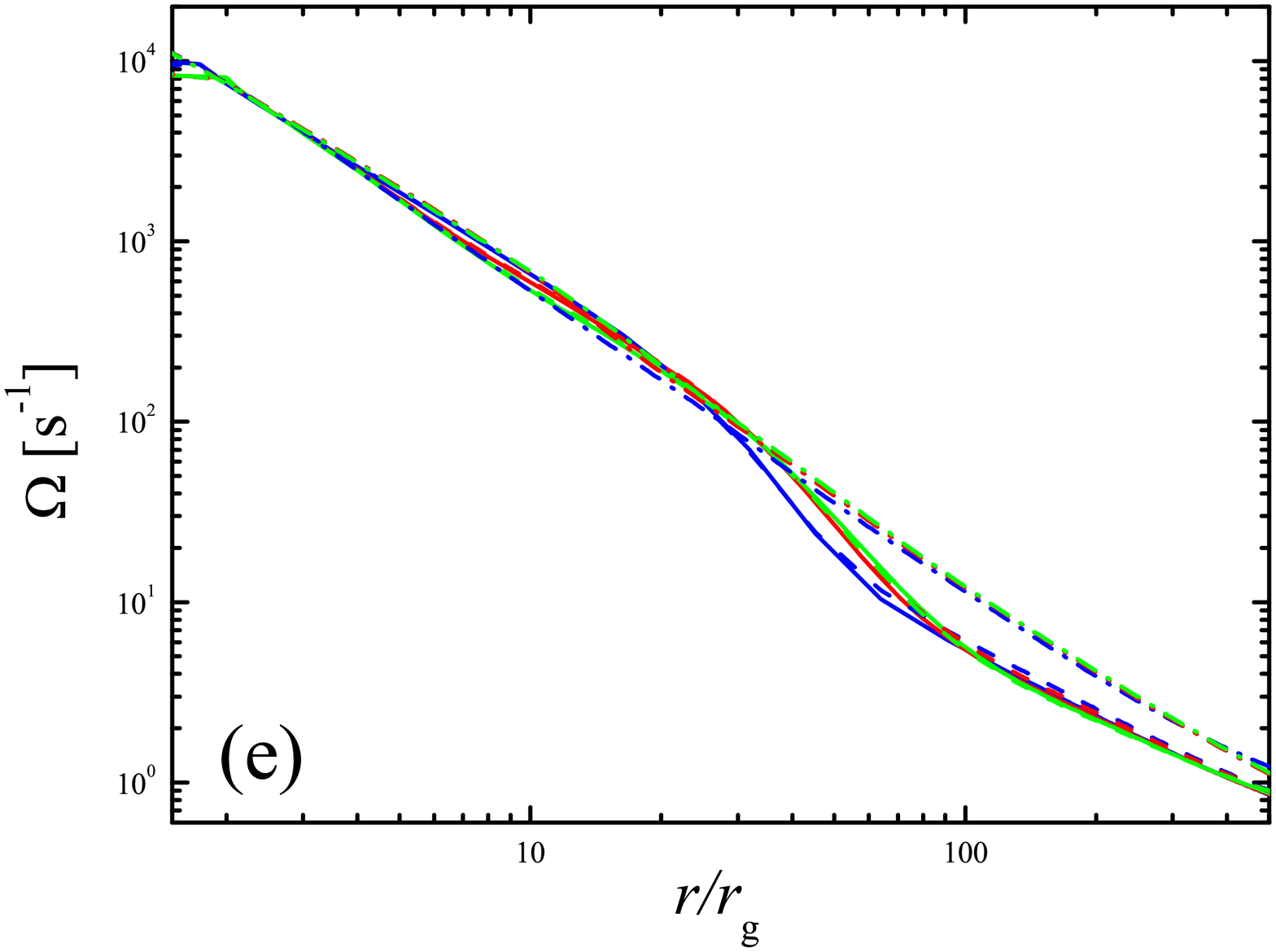}
\includegraphics[width=0.45\textwidth]{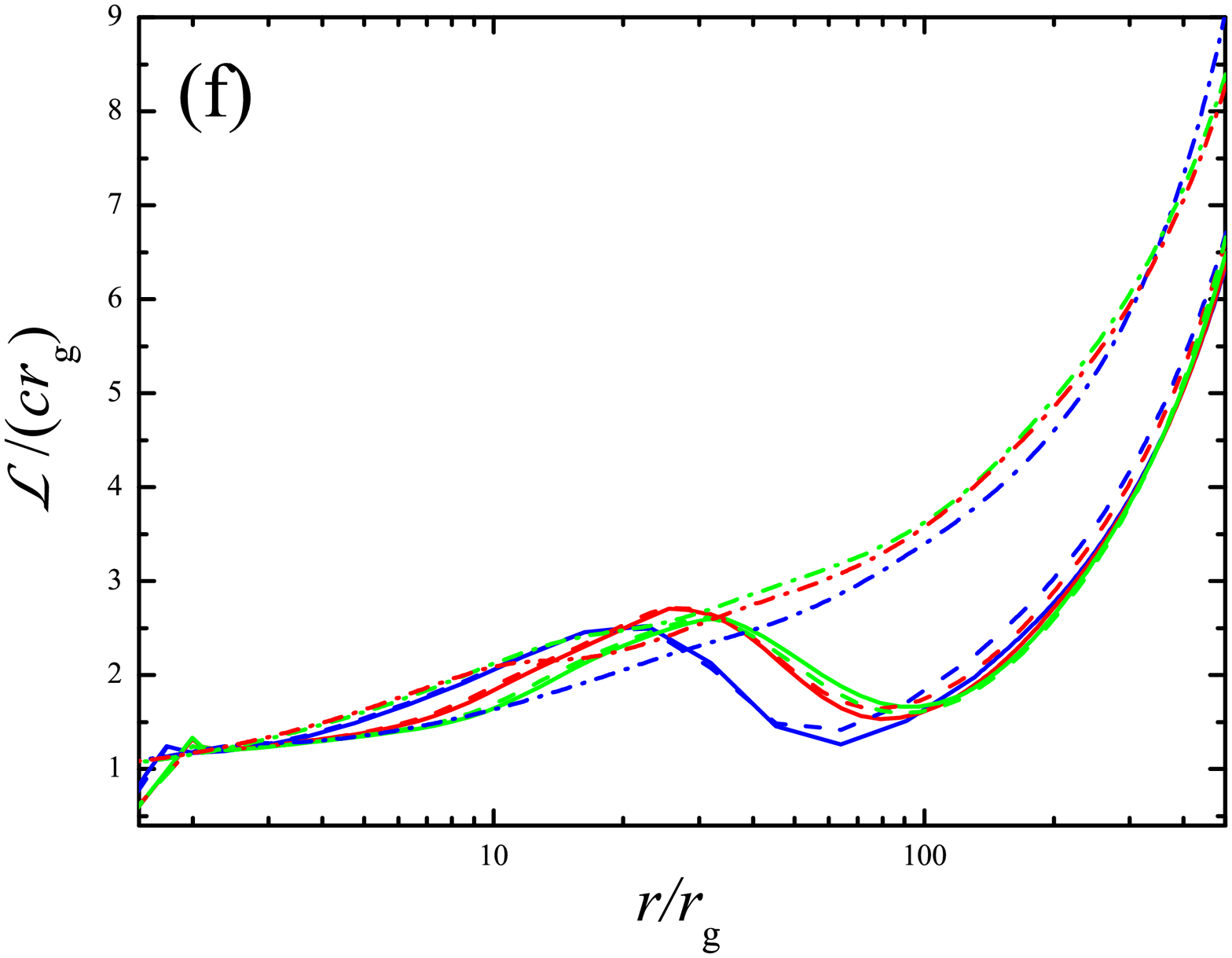}
\includegraphics[width=0.45\textwidth]{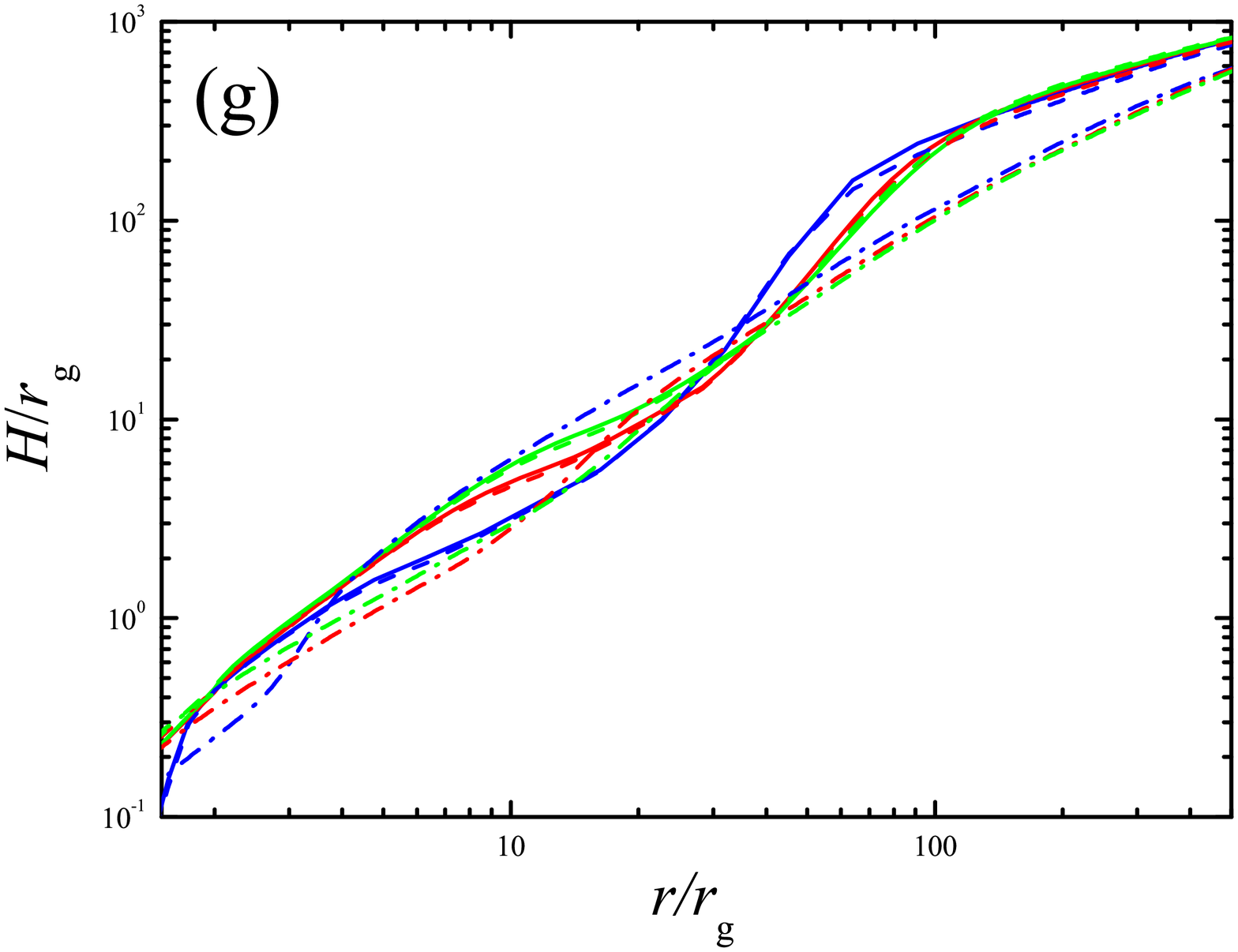}
\caption{Profiles of (a) density $\rho$, (b) temperature $T$, (c) surface density $\Sigma$, (d) absolute value of radial velocity $\mid V_r\mid$, (e) angular velocity $\Omega$, (f) specific angular momentum $\mathcal{L}$ and (g) half-thickness $H$ of MCNDAFs and the counterparts of NDAFs. The blue, red, and green lines correspond to $\dot{m}= 0.1$, 0.5 and 1, respectively. The solid and dashed lines correspond to MCNDAFs with $n=4$ and $5$ as well as the dot-dashed lines denote NDAFs.}
\label{fig_total}
\end{figure*}

\subsubsection{Electron Fraction}

The definition of the electron fraction is given by \citep[e.g.][]{Liu2017b}
\beq
Y_{\rm e}=\frac{\sum\limits_{j} n_j Z_j} {\sum\limits_{j} n_j (Z_j+N_j)},
\eeq
where $Z_j$ and $N_j$ are the number of the protons and neutrons of a nucleus. Here we employ the strict NSE equation (see Section \ref{sec_nucleosynthesis}) to calculate the electron fraction. Meanwhile, the condition of electrical neutrality remain valid, which can be written as \citep[e.g.][]{Liu2007,Liu2013}
\beq
\sum\limits_{j} {n_j} {Z_j} = \frac {\rho Y_{\rm e}}{m_u} = n_{\rm e^{-}}-n_{\rm e^{+}},
\eeq
where $m_u$ is the mean mass of baryons.

Furthermore, the other relation on the number density is a bridging formula. For the extremely neutrino optically thin and thick regions, the chemical potentials are satisfied with $\mu_{\rm e}=\mu_{\rm n}-\mu_{\rm p}$ and $2\mu_{\rm e}=\mu_{\rm n}-\mu_{\rm p}$ \citep[see][]{Yuan2005}, respectively, where $\mu_{\rm n}$ and $\mu_{\rm p}$ are the chemical potentials of neutrons and protons, and the chemical potential of neutrinos is ignored because it is much lower than the chemical potential of electron $\mu_{\rm e}$. Then the ratio of the number density of free neutrons $n_2$ to that of free protons $n_1$ transits from the optically thin to optically thick regimes,
\beq \label{bridging-formula}
\lg{\frac{n_{\rm 2}}{n_{\rm 1}}} = {\rm e}^{-\tau_{\nu_{\rm e}}} \frac{2 \mu_{\rm e}-Q}{k_{\rm B} T}+(1-{\rm e}^{-\tau_{\nu_{\rm e}}}) \frac{\mu_{\rm e}-Q}{k_{\rm B} T},
\eeq
where $Q=(m_{\rm n}-m_{\rm p})c^2$, and $m_{\rm n}$ and $m_{\rm p}$ are the masses of the neutron and proton, respectively \citep[e.g.][]{Liu2007,Xue2013,Liu2017b}. For the more detailed descriptions on the modification of the $\beta$-equilibrium condition from neutrino optically thin to thick, see \citet{Li2013}.

\subsection{Nucleosynthesis}\label{sec_nucleosynthesis}

NSE is established when all nuclear reactions achieve chemical equilibrium. We adopt the NSE state proposed by \citet{Seitenzahl2008}, which is satisfied with the entire range of the electron fraction\footnote{\url{http://cococubed.asu.edu/code_pages/nse.shtml}}. The number density of nucleus $n_j$ can be expressed as
\beq
n_j =&& g_j (\frac{m_j k_{\rm B} T}{\hbar^2})^{3/2}\nonumber\\ &&\times {\rm exp}[\frac{Z_j (\mu^{\rm kin}_{\rm p}+{\mu^{\rm C}_{\rm p}})+N_j {\mu^{\rm kin}_{\rm n}}-{\mu^{\rm C}_{\rm j}}+Q_j}{k_{\rm B} T}],\nonumber\\
\eeq
where $\mu^{\rm kin}_{\rm p}$ and $\mu^{\rm kin}_{\rm n}$ are the kinetic chemical potentials of protons and neutrons, $\mu^{\rm C}_{\rm p}$ and $\mu^{\rm C}_{\rm j}$ are the Coulomb chemical potentials of protons and nucleons, and $g_j$ is the nuclear partition functions.

\subsection{Thermodynamics}\label{sec_thermodynamics}

The equation of state in our model can be written as
\beq
p=p_{\rm gas}+p_{\rm rad}+p_{\rm e}+p_\nu+p_{\rm mag}.
\eeq

The gas pressure from free nucleons $p_{\rm gas}$ is estimated by
\beq
p_{\rm gas}= \sum\limits_{j} n_j k_{\rm B} T.
\eeq

The photon radiation pressure $p_{\rm rad}$ is
\beq
p_{\rm rad}=a_{\rm rad} T^4/3,
\eeq
which form is caused by the totally optical thick for photon in the disc, and $a_{\rm rad}$ is the radiation constant.

The electron pressure $p_{\rm e}$ is composed of the contributions from electrons and positrons. Since there is no asymptotic expansion valid for electrons with moderate degeneracy and relativity, it has to be calculated by the exact Fermi-Dirac distribution \citep[e.g.][]{Chen2007,Liu2007,Liu2017b}. It reads
\beq
p_{\rm e}=p_{\rm e^{-}}+ p_{\rm e^{+}},
\eeq
with
\beq
&&p_{\rm e^{\mp}}= \frac{1}{3 {\pi}^2 {\hbar}^3 c^3} \nonumber\\&& \times \int_0^{\infty} d p \frac{p^4}{\sqrt{p^2 c^2+{m_{\rm e}}^2 c^4}} \frac{1}{{\rm e}^{({\sqrt{p^2 c^2+{m_{\rm e}}^2 c^4} \mp {\mu_{\rm e}})/k_{\rm B} T}}+1}.\nonumber\\
\eeq

The neutrino pressure $p_\nu$ is
\beq
p_\nu=u_\nu/3,
\eeq
where ${u}_{\nu}$ is the energy density of neutrinos, for which we adopt the bridging formula \citep[e.g.][]{Kohri2005,Liu2007,Liu2017b}, i.e.,
\beq \label{neutrino_density}
u_\nu=\sum_{i} \frac{(7/8)a_{\rm rad} T^4 (\tau_{{\nu}_i}/2+1/ \sqrt{3})} {\tau_{{\nu}_i}/2+1/ \sqrt{3}+1/(3 \tau_{a,{\nu}_i})}.
\eeq
It is valid in the regimes from the optically thin to thick for neutrinos.

The magnetic pressure contributed by the tangled magnetic fields in the disc is
\beq
p_{\rm mag} =\beta_t p,
\eeq
and $\beta_t$ is the ratio of the magnetic pressure to the total pressure, we adopt $\beta_t=0.1$ in our calculation.

The cooling rate $Q^-$ appearing in Equation (\ref{energy}) can be defined as
\beq
Q^{-}=Q_{\rm ph} +Q_{\nu}+Q_{\rm rad},
\eeq
which consists of the cooling rates of photodisintegration $Q_{\rm ph}$, the neutrino emitting $Q_{\nu}$ and the photon radiation $Q_{\rm rad}$. However, due to photon trapping by dense material, $Q_{\rm {rad}}$ is always much smaller than the other two, so we ignored this term in our calculation \citep[e.g.][]{Liu2017b}.

The cooling rate of photodisintegration $Q_{\rm ph}$ is mainly contributed by the decomposition of $\alpha$-particles, which can be written as
\beq
Q_{\rm ph} = 6.8 \times 10^{28} \rho_{10} V_r H \frac{d X_{\rm nuc}}{d r}~{(\mathrm {cgs~units})},
\eeq
where ${\rho}_{10} \equiv {\rho}/(10^{10}{\rm g~cm^{-3}})$, and $X_{\rm nuc}$ is the mass fraction of $\alpha$-particles \citep[e.g.][]{Kohri2005,Liu2007,Liu2017b}.

According to Equation (\ref{neutrino_density}), the cooling rate due to the neutrino loss $Q_\nu$ can be obtain,
\beq
Q_{\nu}=\sum_{i} \frac{(7/8) {\sigma} T^4}{(3/4)[\tau_{{\nu}_i}/2+1/ \sqrt{3}+1/(3 \tau_{a,{\nu}_i})]}.
\eeq

\begin{figure*}
\centering
\includegraphics[width=0.48\textwidth]{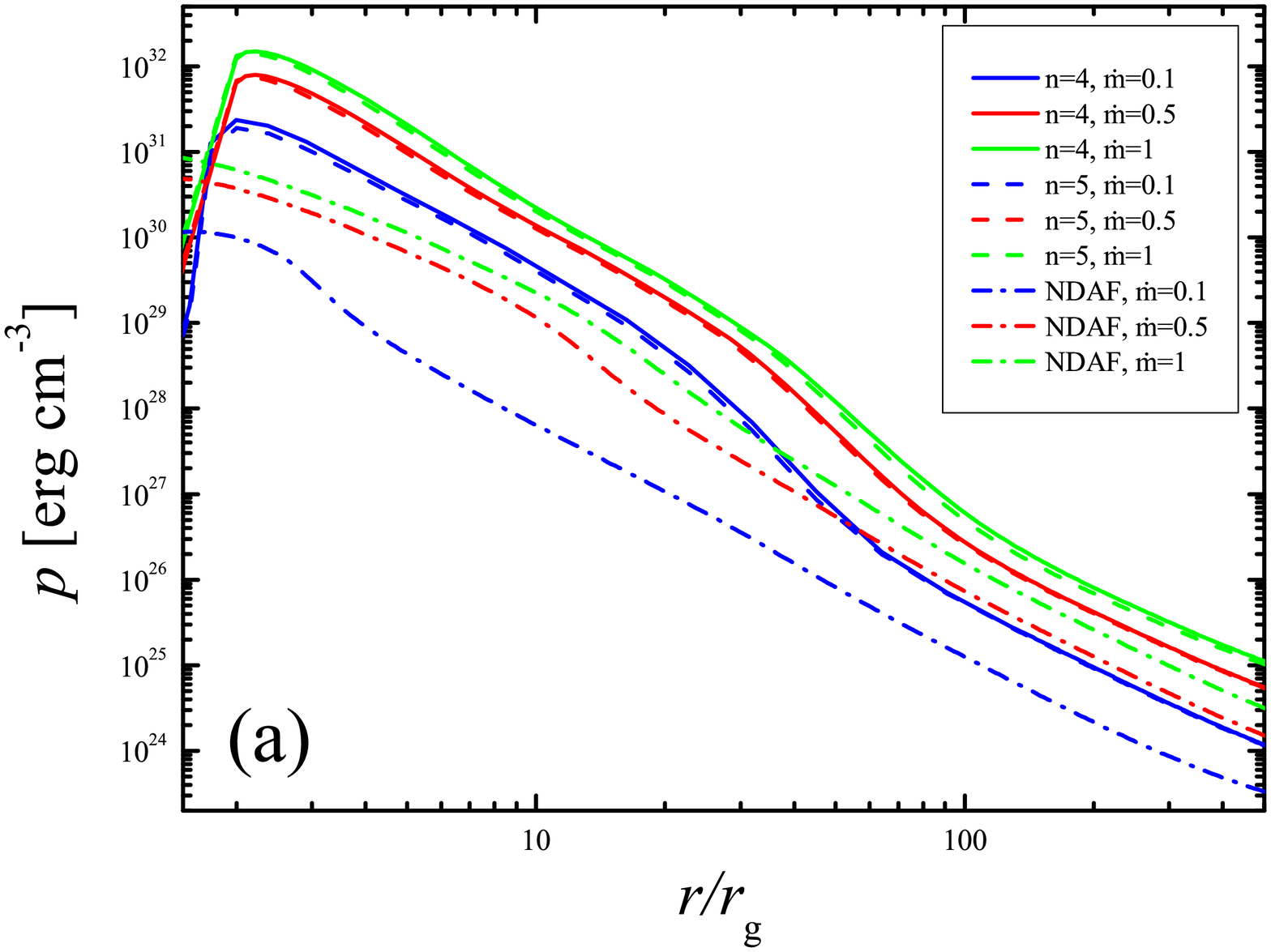}
\includegraphics[width=0.48\textwidth]{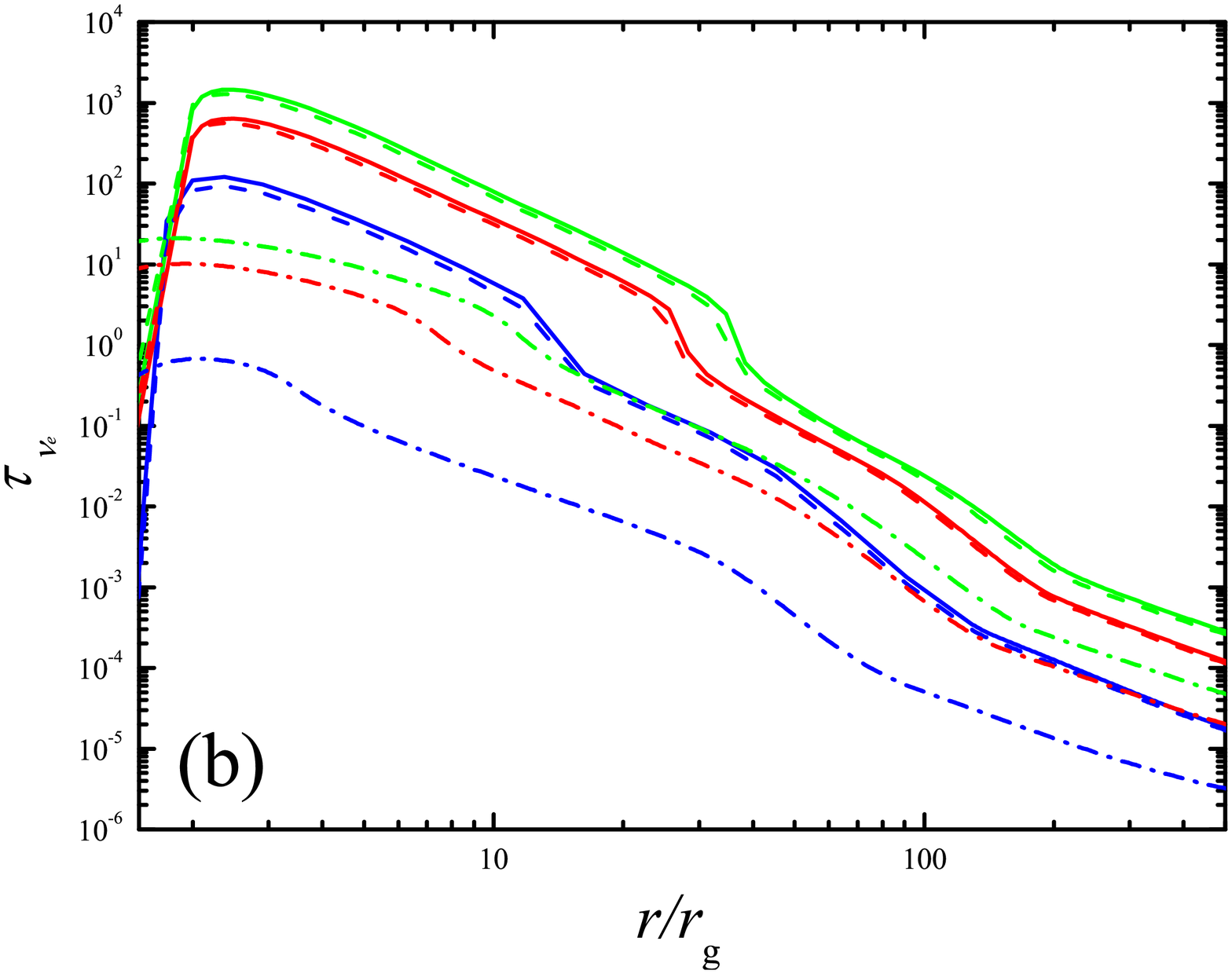}
\includegraphics[width=0.48\textwidth]{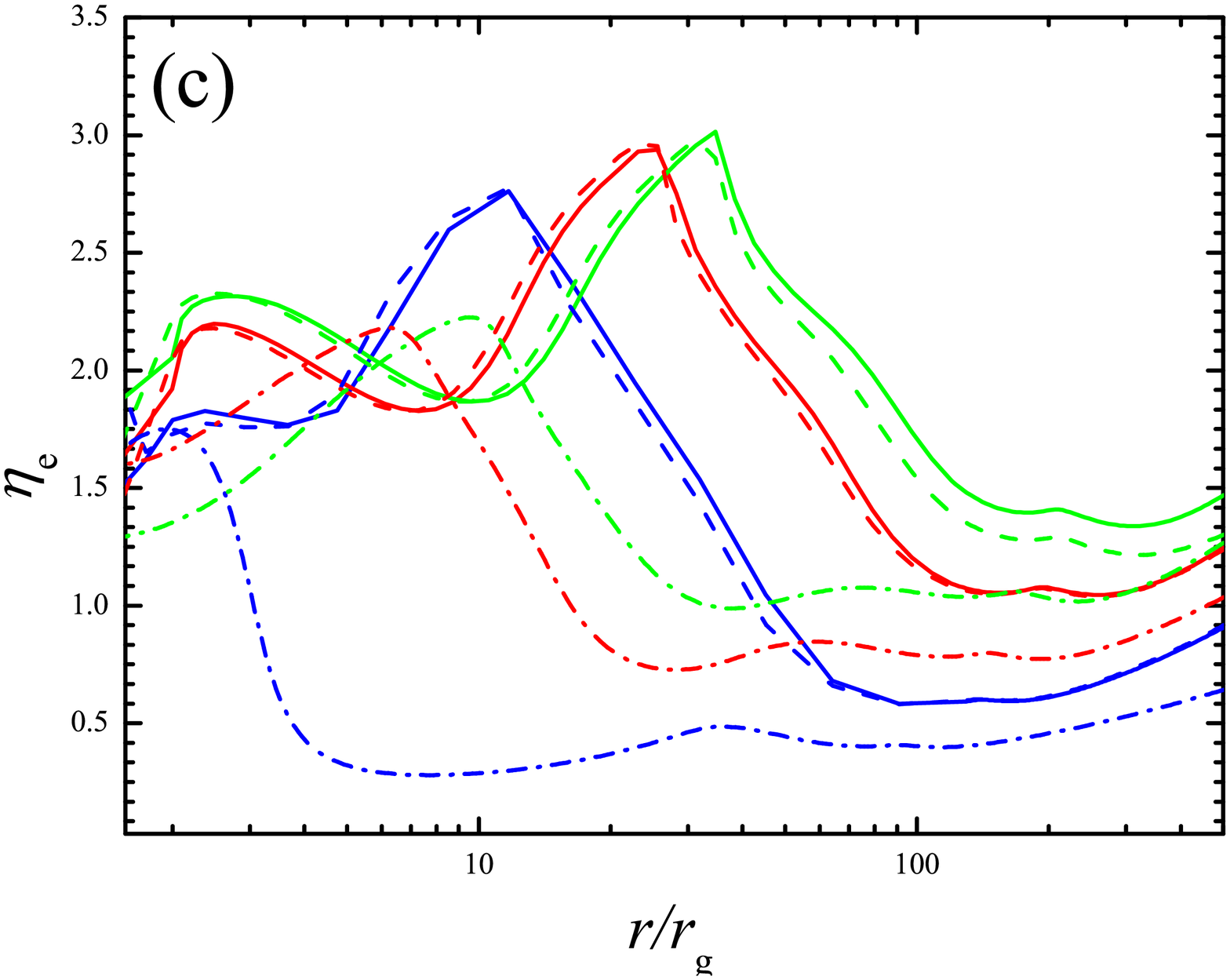}
\includegraphics[width=0.48\textwidth]{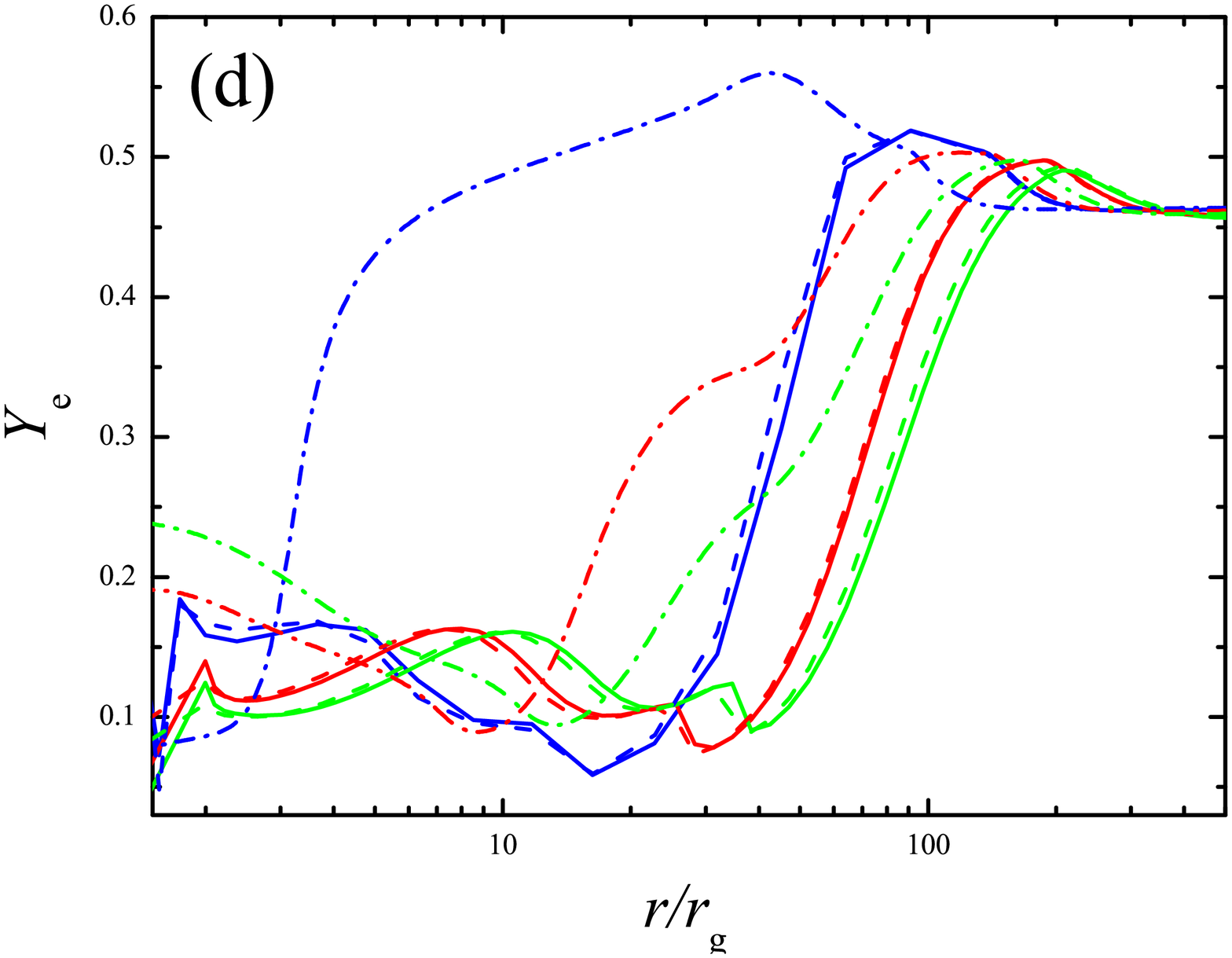}
\caption{Profiles of (a) total pressure $p$, (b) optical depth of electron neutrinos $\tau_{\nu_e}$, (c) electron degeneracy $\eta_e$, and (d) electron fraction $Y_{\rm e}$ with the same denotations of colours and line-styles as Figure 1.}
\label{fig_total1}
\end{figure*}

\begin{figure*}
\centering
\includegraphics[width=0.48\linewidth]{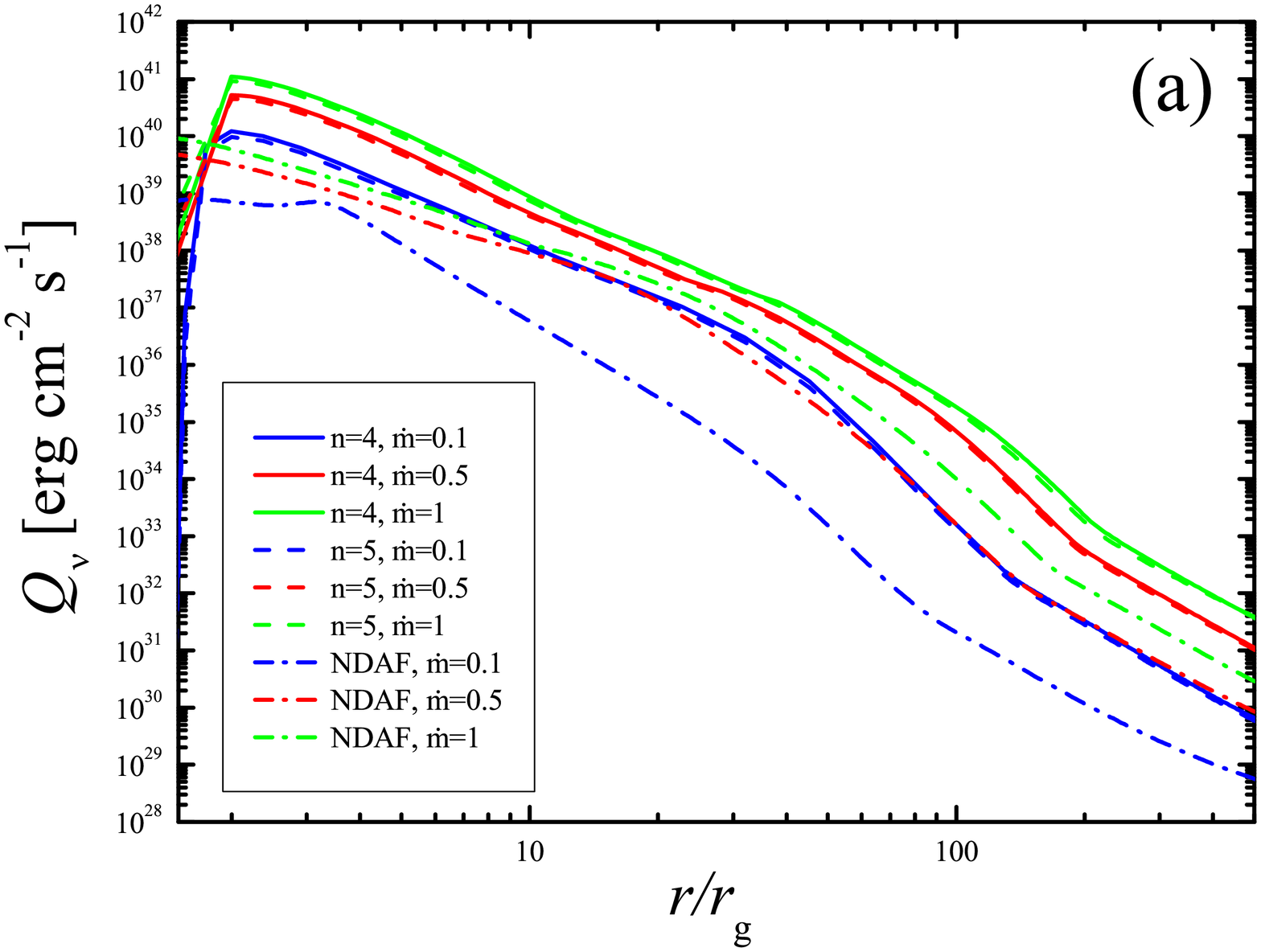}
\includegraphics[width=0.48\linewidth]{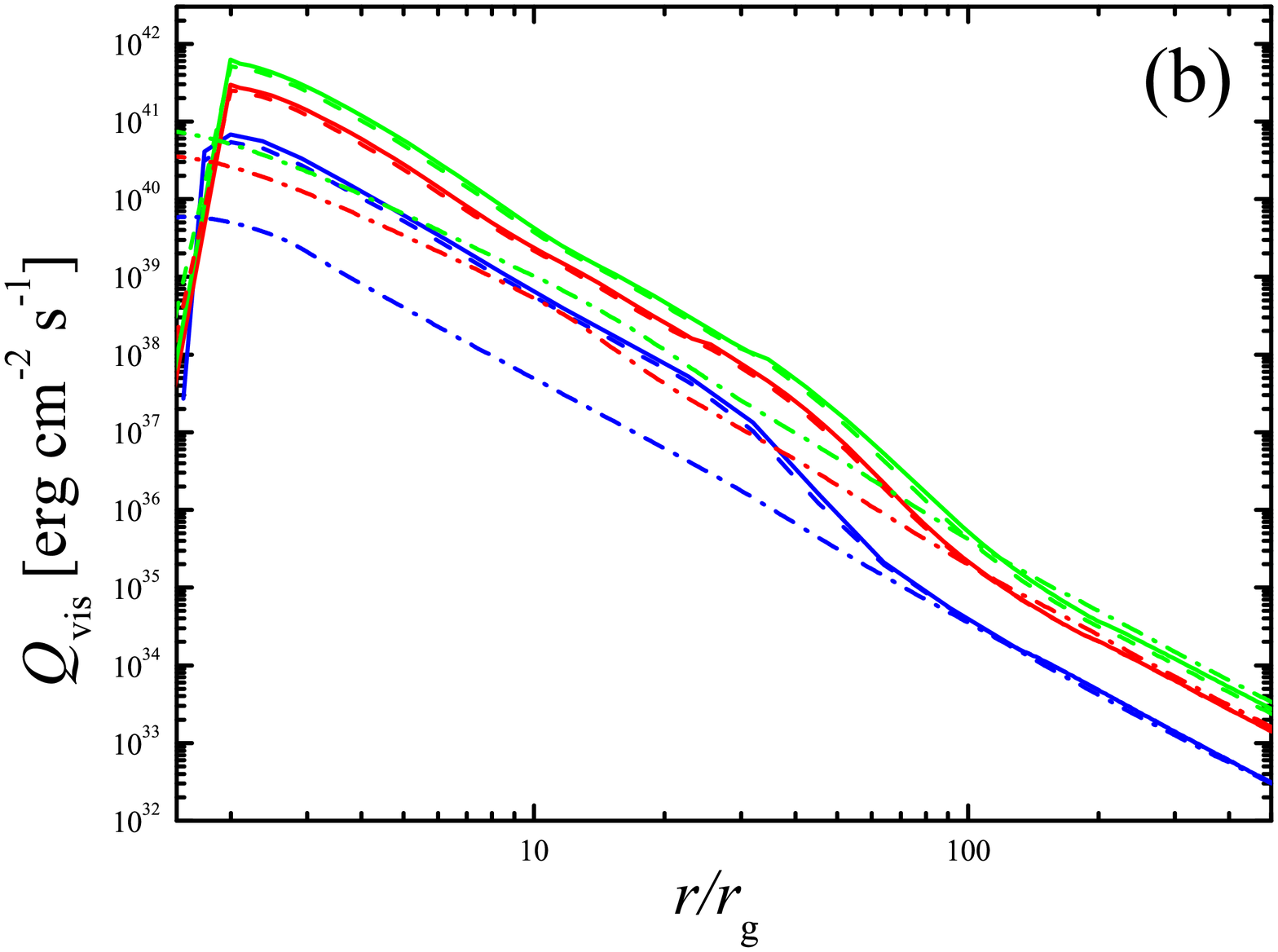}
\includegraphics[width=0.48\linewidth]{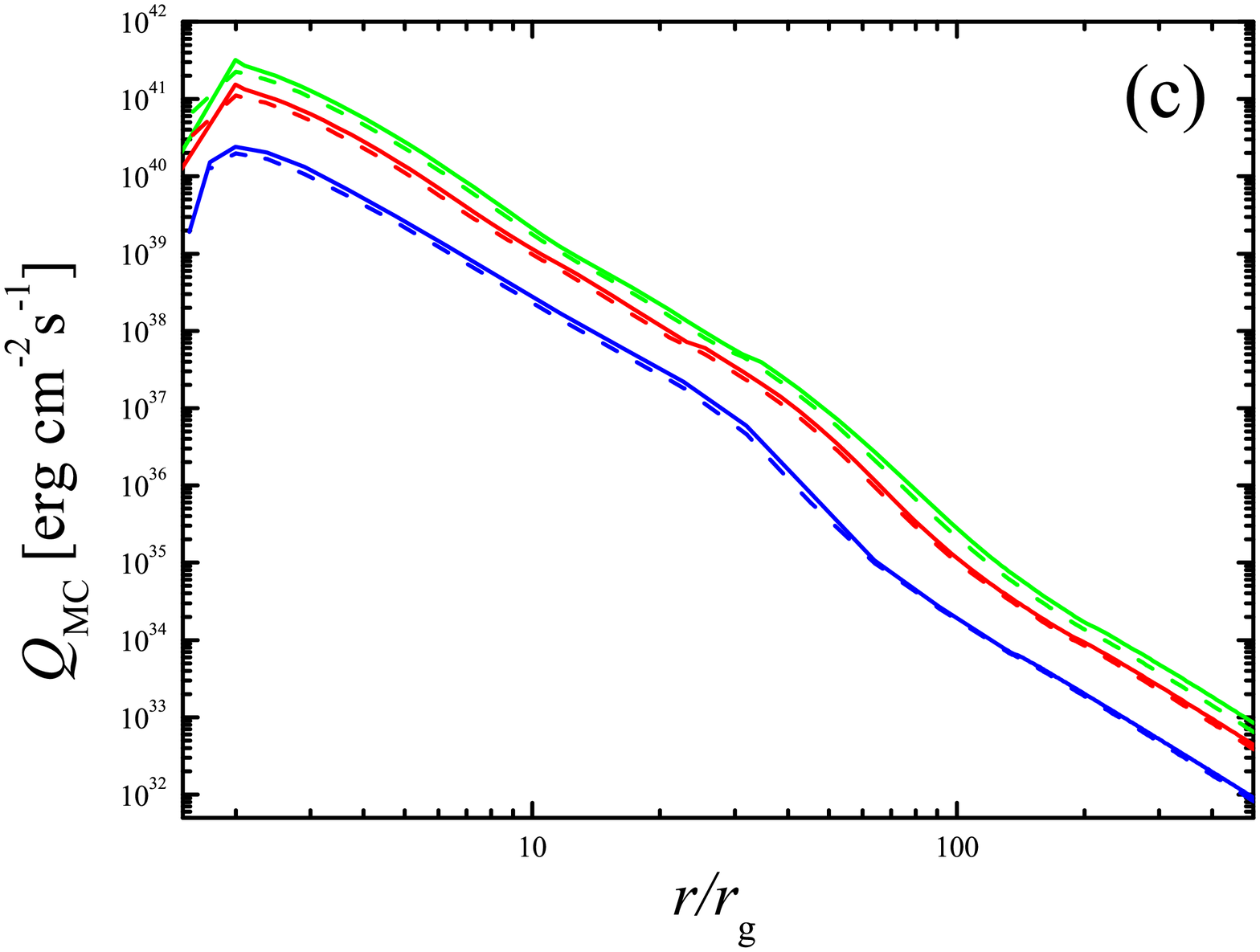}
\includegraphics[width=0.48\linewidth]{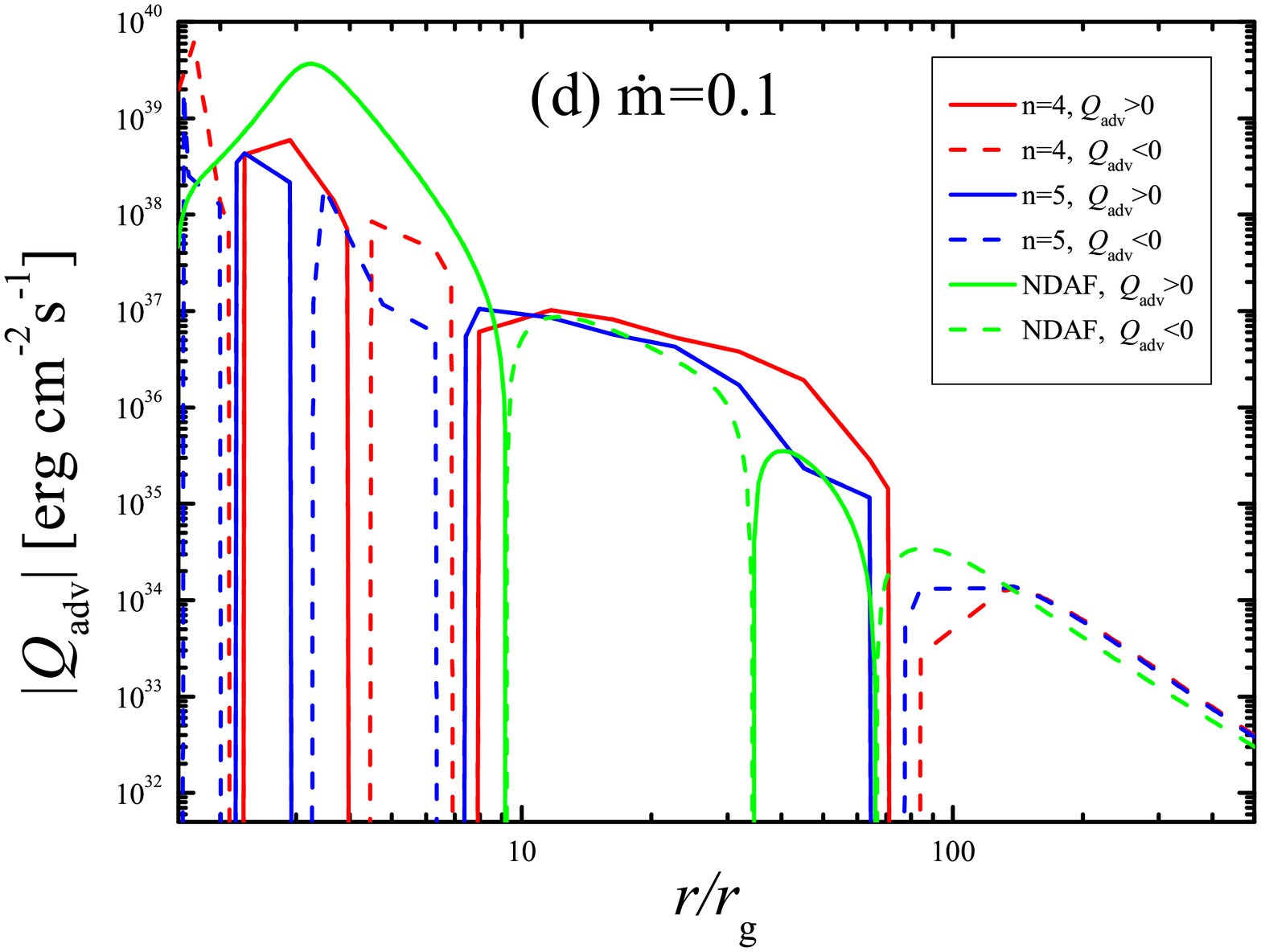}
\includegraphics[width=0.48\linewidth]{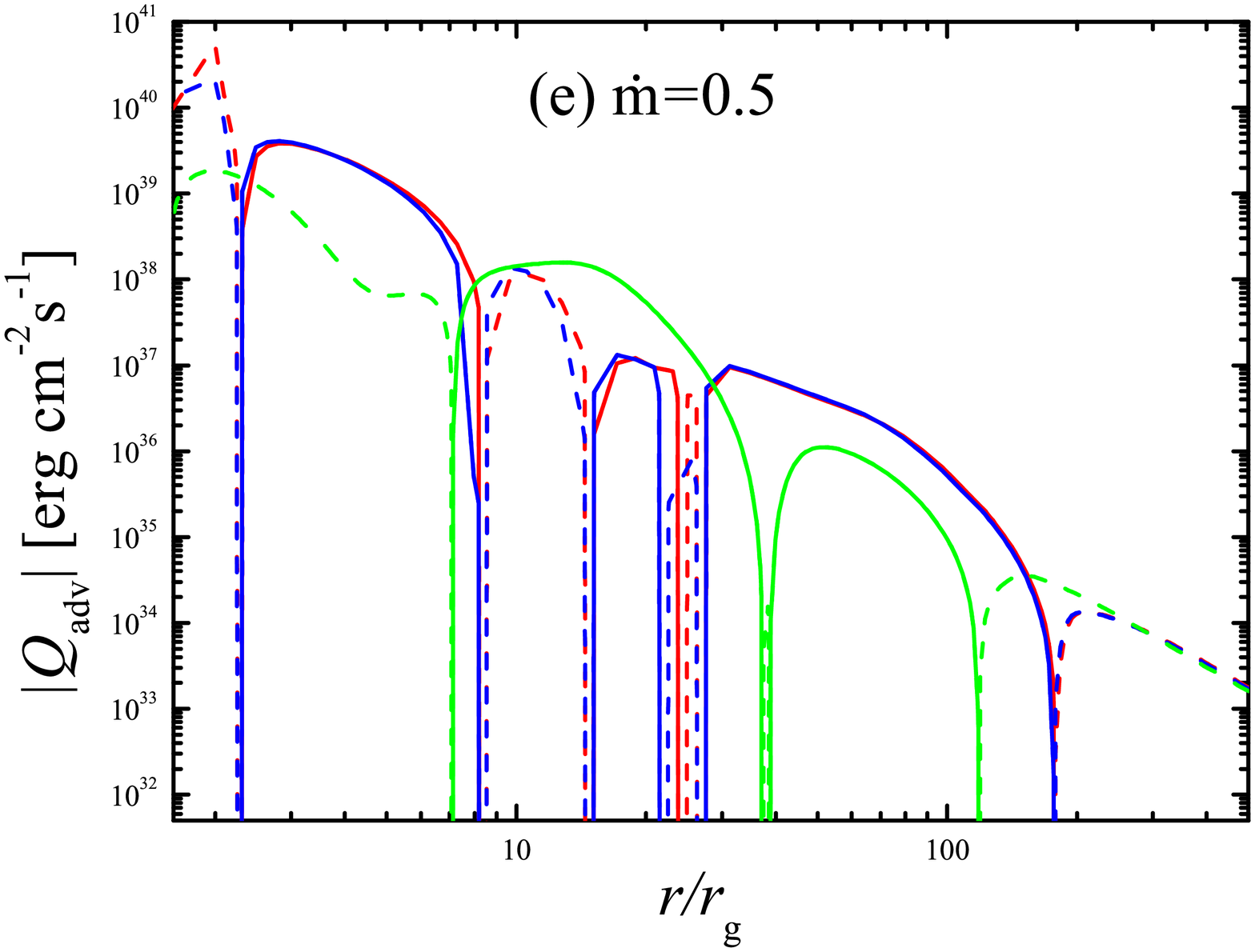}
\includegraphics[width=0.48\linewidth]{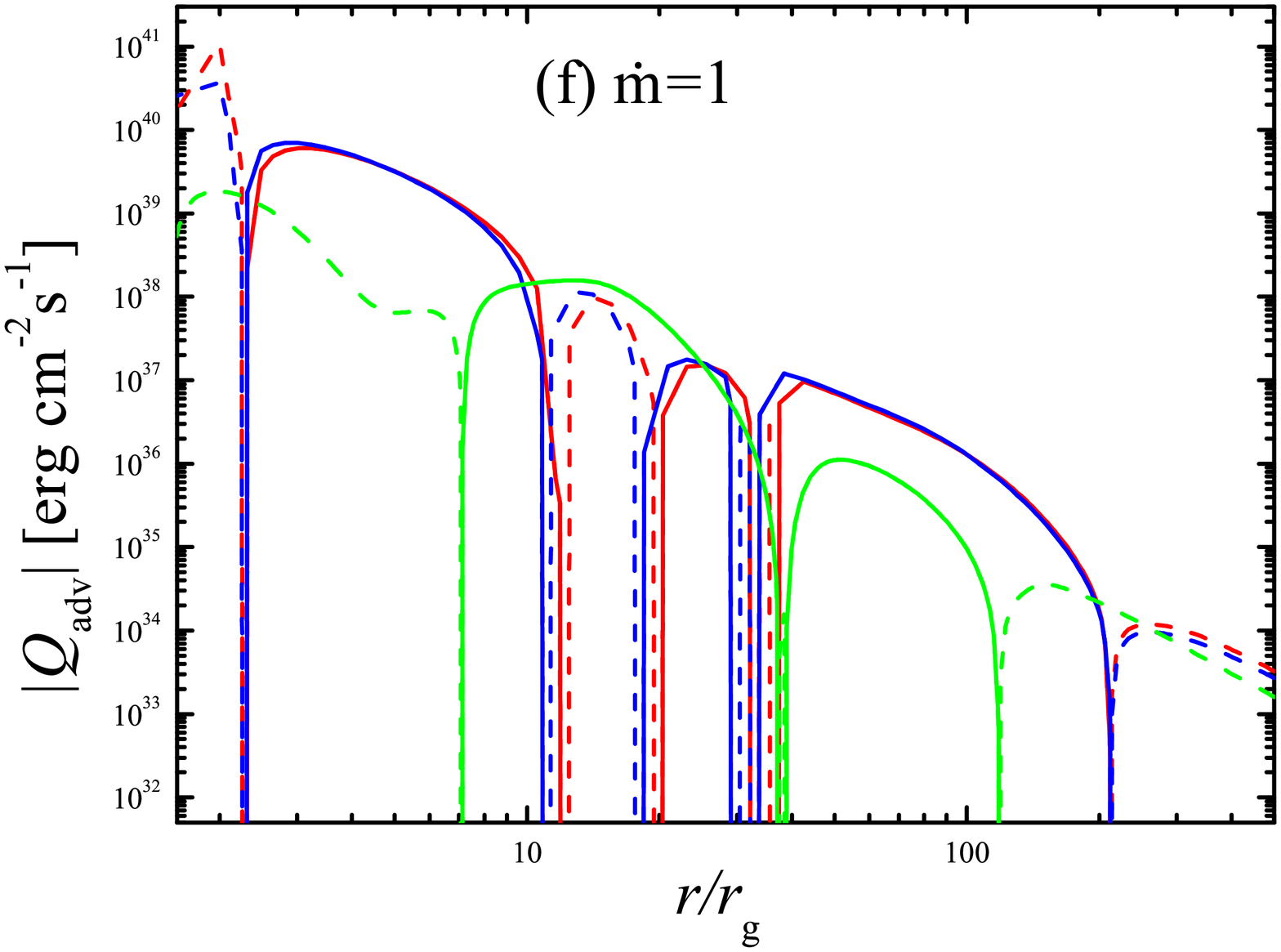}
\caption{Radial distributions of (a) neutrino cooling rates $Q_{\nu}$, (b) viscous heating rates $Q_{\rm{vis}}$, (c) MC heating rates $Q_{\rm{MC}}$, and (d-f) absolute values of advection cooling rates $\mid Q_{\rm {adv}}\mid$ for $\dot{m}= 0.1$, 0.5 and 1. \emph{Panels} (a-c): the blue, red, and green lines correspond to $\dot{m}=$ 0.1, 0.5 and 1, respectively, and the solid and dashed lines correspond to MCNDAFs with $n=4$ and $5$ as well as the dot-dashed lines denote NDAFs. \emph{Panels} (d-f): the red and blue lines correspond to MCNDAFs with $n=4$ and $5$ as well as the green lines denote NDAFs. The solid and dashed parts of each line denote the positive and negative values of $Q_{\rm{adv}}$.}
\label{fig_energy1}
\end{figure*}

The MC torque $T_{\rm MC}$ and the energy transferred from the BH to the disc by the MC process $Q_{\rm MC}$ are given by \citep[e.g.][]{Wang2002,Lei2009,Luo2013,Song2020},
\beq
T_{\rm MC}=4T_0 a_*(1+q)\int ^{\pi/2}_{\theta_0} \frac{(1-\Omega/\Omega_{\rm H}) \sin^3\theta d\theta}{2-(1-q)\sin^2\theta},
\eeq
and
\beq
Q_{\rm MC}=\frac{T_{\rm MC}}{4\pi r}  \frac{d\Omega}{dr},
\eeq
respectively, where $q=a_*/(1+\sqrt{1-a_*^2})$ is a dimensionless parameter relating to the dimensionless BH spin $a_*$, $T_0=3.26\times10^{45}(B_{\rm H}/10^{15}~{\rm G})^2(M_{\rm BH}/M_\odot)^3 ~\rm{g~cm^2~s^{-2}}$ is a characteristic torque, $\Omega_{\rm H}=(a_*c/2r_{\rm g})(1+q)$ is the angular velocity at the BH horizon. $B_{\rm H}^2=8\pi c \dot{M}/r_{\rm g}^2$ is the field strength at the BH horizon \citep[e.g.][]{McKinney2005,Lei2009}, where $r_{\rm g}=2G M_{\rm BH}/c^2$ is the Schwarzschild radius, while magnetic field of inner region vary as $B_{\rm D}\propto \xi^{-n}$ described in \citet{Blandford1976}, where $\xi=r/r_{\rm ms}$, $r_{\rm ms}$ is the marginally stable orbital radius, and $n$ is the index measuring the concentration of the magnetic fields on the equatorial plane. Moreover, $\theta_0$ is the azimuth corresponding to the boundary between the open and closed field lines, which will tend to zero when the MC effect vanishes and can be defined as \citep[e.g.][]{Wang2002}
\beq
&&\cos\theta_0 = \int^{\xi_{\rm out}}_1 d\xi \nonumber\\ && \times \frac{\xi^{1-n}\chi_{\rm ms}^2\sqrt{1+a_*^2\chi^{-4}_{\rm ms}\xi^{-2}+2a_*^2\chi^{-6}_{\rm ms}\xi^{-3}}}
{2\sqrt{(1+a_*^2\chi^{-4}_{\rm ms}+2a^2_*\chi^{-6}_{\rm ms})(1-2\chi^{-2}_{\rm ms}\xi^{-1}+a^2_*\chi^{-4}_{\rm ms}\xi^{-2})}},\nonumber\\
\eeq
where $\chi_{\rm ms}=\sqrt{r_{\rm ms}/r_{\rm g}}$ is a dimensional parameter relating to the marginally stable orbital radius and $\xi_{\rm out}$ is the the dimensionless outer boundary of the MC dominated region.

\subsection{Neutrino and annihilation luminosities}

\begin{figure*}
\centering
\includegraphics[width=0.47\linewidth]{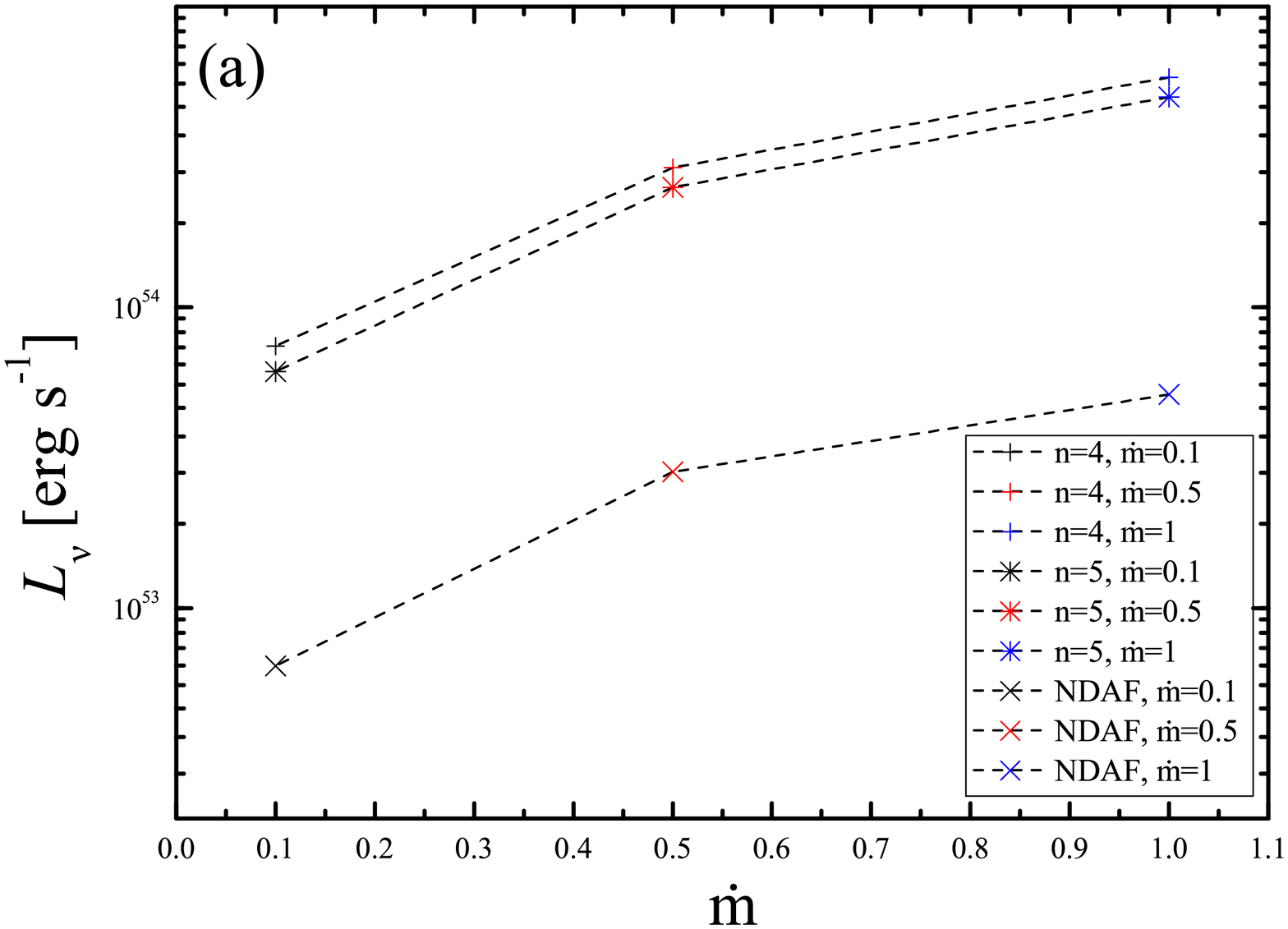}
\includegraphics[width=0.47\linewidth]{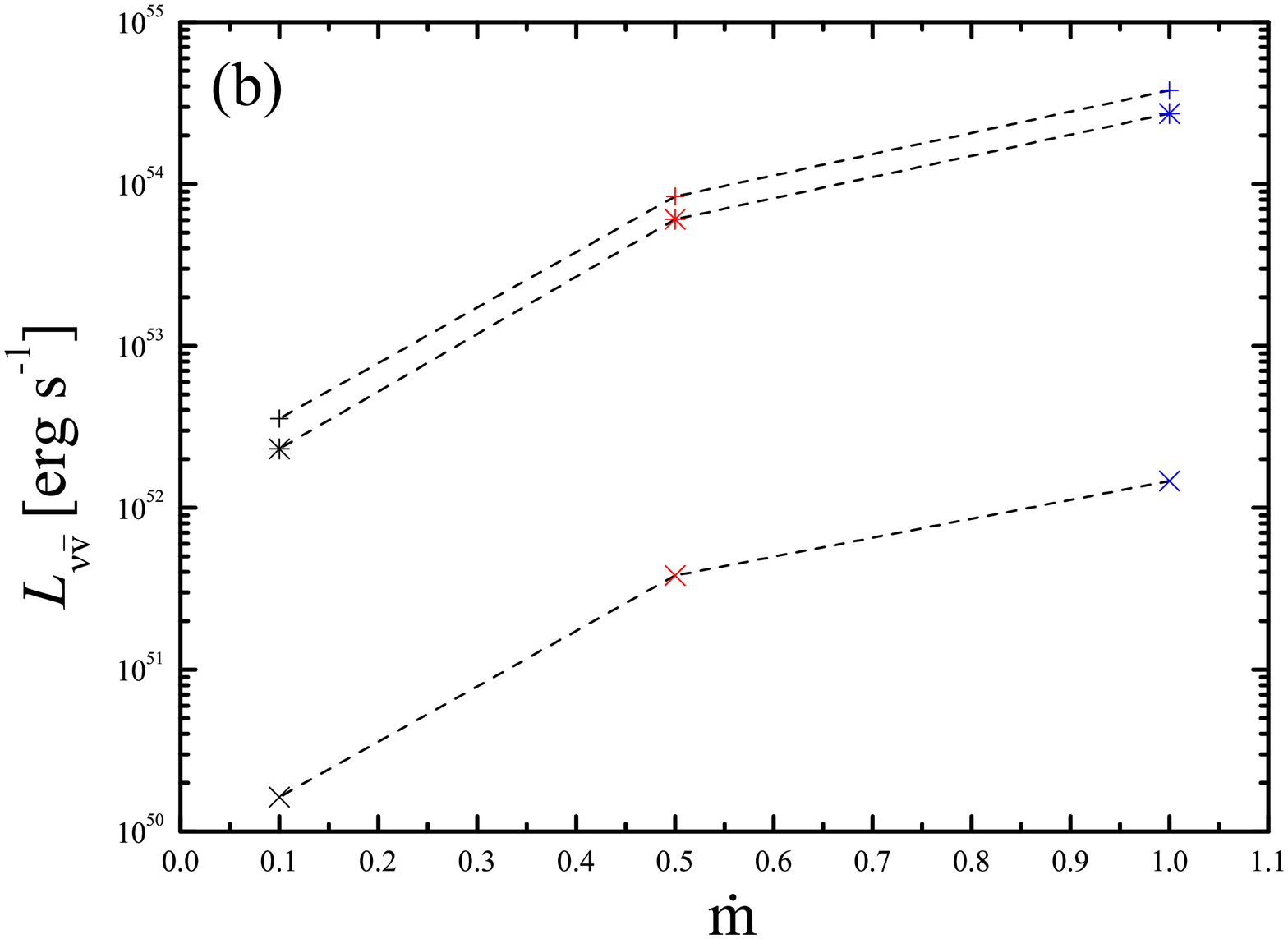}
\caption{(a) Neutrino luminosity and (b) neutrino annihilation luminosity as functions of the accretion rates. The black, red, and blue markers correspond to $\dot{m}$ = 0.1, 0.5, and 1, respectively. The pluses and asterisks correspond to MCNDAFs with $n=4$ and $5$ as well as the crosses denote NDAFs.}
\label{fig_lvlvv}
\end{figure*}

Due to the neutrino physics in MCNDAFs, the electron neutrinos and antineutrinos are the main products to annihilate in the space out of the disc \citep[e.g.][]{Chen2007,Xue2013,Liu2017b}. After obtaining the radial distribution of the neutrino cooling rate $Q_{\nu}$, the neutrino luminosity can be calculated by integrating along radial direction, i.e.,
\beq
L_\nu=4 \pi \int_{{\rm max}(r_{\rm ms}, r_{\rm tr})}^{r_{\rm out}} Q_\nu r d r,
\eeq
where the outer boundary of the disc $r_{\rm out}=500~r_{\rm g}$ is set in our calculations. The inner edge of integration is set to be the larger one of the neutrino trapping radius $r_{\rm tr}$ and $r_{\rm ms}$ to take into account of the effect of neutrino trapping.

The neutrino trapping radius can be obtained by comparing the neutrino diffusion timescale $t_{\rm diff}$ with the accretion timescale $t_{\rm acc}$. In other words, the criterion should be given to judge that the neutrinos produced in the equatorial planes of the disc can escape from the disc surface instead of falling into the BH. The photon diffusion timescale in classical accretion discs can be estimated as $\sim H / (c/3 \tau)$ \citep[e.g.][]{Ohsuga2002,Liu2017b}. Thus, for MCNDAFs, we estimate $t_{\rm diff} \sim H/(V_{\rm n}/3\tau_{\nu_{\rm e}})$, where the neutrino speed $V_{\rm n}\sim [1-(0.07{\rm eV} /3.7 k_{\rm B} T)]^{1/2}~c$ \citep[e.g.][]{Liu2012b,Liu2017b} and the photon optical depth $\tau$ is substituted by the neutrino optical depth $\tau_{\nu_{\rm e}}$. The other one, $t_{\rm acc}\sim -r/V_r$. The lower limit of neutrino mass $0.07~{\rm eV}$ is adopted here. According to the condition $t_{\rm diff}>t_{\rm acc}$, the neutrino trapping radius can be roughly estimated by
\beq
r_{\rm tr} \approx -\frac{3 \tau_{\nu_{\rm e}}H V_r}{V_{\rm n}}.
\eeq
Obviously, the effect of neutrino trapping will greatly affect the annihilation luminosity if it exists generally for the high accretion rates.

We adopted the Newtonian approximation method \citep[e.g.][]{Ruffert1997,Popham1999,Rosswog2003,Liu2007,Liu2017b,Xue2013} to calculate the neutrino annihilation luminosity, the value of which has the same order of magnitude to the relativity method \citep{Zalamea2011}. In this approach, the disc is modeled as a grid of cells in the equatorial plane. Every cell $k$ has its mean neutrino energy $\varepsilon_{\nu_i}^k$, neutrino radiation luminosity $l_{\nu_i}^k$, and distance to a space point in the outside space of the disc $d_k$. $l_{\nu_i}^k$ can be calculated by the surface integral of cooling rate of each flavor of neutrino in the cell $k$. The angle at which neutrinos from cell $k$ encounter antineutrinos from another cell $k'$ at that point is denoted as $\theta_{kk'}$. Then the neutrino annihilation luminosity at that point is described by the summation over all pairs of cells,
\beq
l_{\nu \overline{\nu}}=\sum_{i} A_{1,i} \sum_k \frac{l_{\nu_i}^k}{d_k^2} \sum_{k'} \frac{l_{\overline{\nu}_i}^{k'}}{d_{k'}^2}(\varepsilon_{\nu_i}^k + \varepsilon_{\overline{\nu}_i}^{k'}) {(1-\cos {\theta_{kk'}})}^2 \nonumber\\+ \sum_{i} A_{2,i} \sum_k \frac{l_{\nu_i}^k}{d_k^2} \sum_{k'} \frac{l_{\overline{\nu}_i}^{k'}}{d_{k'}^2} \frac{\varepsilon_{\nu_i}^k +\varepsilon_{\overline{\nu}_i}^{k'}}{\varepsilon_{\nu_i}^k \varepsilon_{\overline{\nu}_i}^{k'}} {(1-\cos {\theta_{kk'}})},\nonumber\\
\eeq
with $A_{1,i} = (1 / 12\pi^2)[\sigma_0/c{(m_{\rm e} c^2)}^2][{(C_{V,\nu_{i}}-C_{A,\nu_{i}})}^2 +{(C_{V,\nu_{i}}+C_{A,\nu_{i}})}^2]$ and $A_{2,i} = (1/6\pi^2) (\sigma_0/c)$ $(2 C_{V,\nu_{i}}^2-C_{A,\nu_{i}}^2)$. The parameters $C_{V,\nu_{i}}$ and $C_{A,\nu_{i}}$ can be found in \citet{Liu2017b}, which are included in the descriptions of the scattering and absorption of neutrinos.

Then the total annihilation luminosity can be obtained by the integration over the whole space outside of the BH and the disc,
\beq
L_{\nu \overline{\nu}}=4 \pi \int_{{\rm max}(r_{\rm ms}, r_{\rm tr})}^\infty \int_H^\infty l_{\nu \overline{\nu}} r d r d z,
\eeq
where the inner edge is depended on the status of neutrino trapping \citep[e.g.][]{Xue2013,Liu2017b}.

\begin{figure*}
\centering
\includegraphics[width=0.48\linewidth]{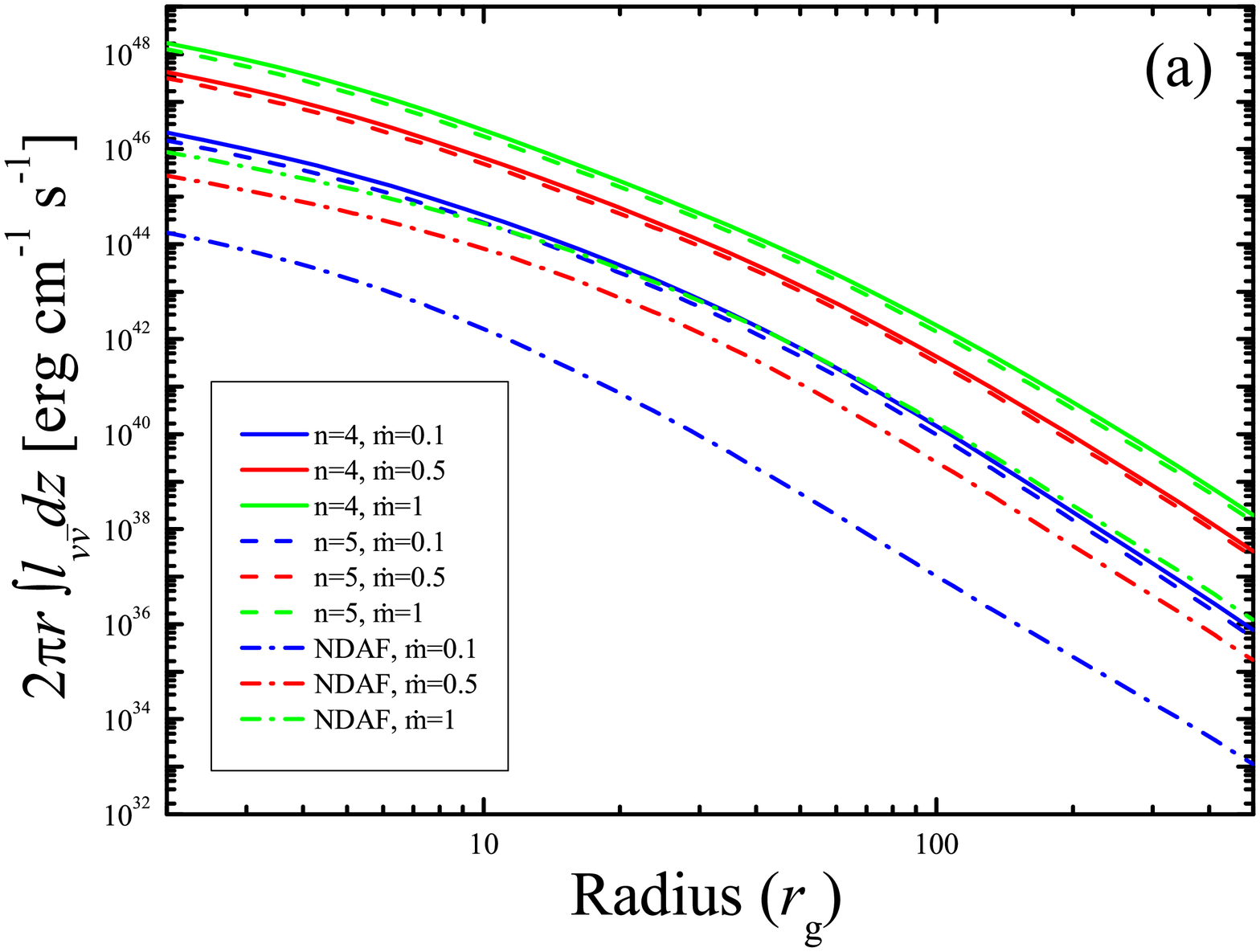}
\includegraphics[width=0.48\linewidth]{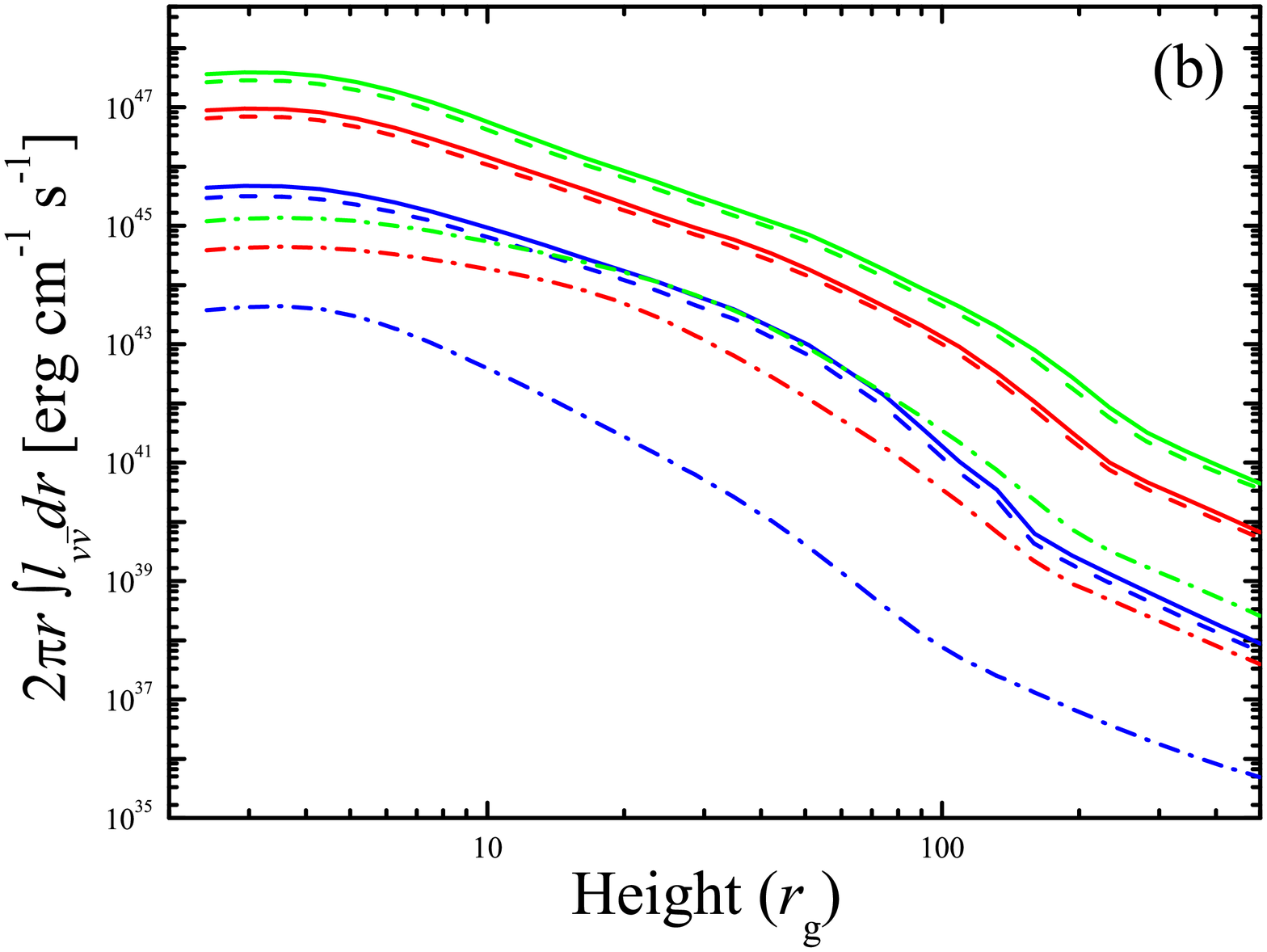}
\caption{Pair-annihilation luminosities per $r_{\rm g}$ (a) vs. radius and (b) vs. distance along the angular momentum axis. The blue, red, and green lines correspond to $\dot{m}$ = 0.1, 0.5 and 1, respectively, and the solid and dashed lines correspond to MCNDAFs with $n=4$ and $5$ as well as the dot-dashed lines denote NDAFs.}
\label{fig_lvvrz}
\end{figure*}
\section{Results}\label{sec_num_res}

In this paper, we use a shooting method \citep{Matsumoto1984} for the solving of above equations, which has been proved to be effective in dealing with the boundary value problem of NDAFs by our previous works \citep[e.g.][]{Xue2013,Liu2017b}. Of course, the convergence of numerical calculations in the MCNDAF model is harder than that in the NDAF model, especially for the inner disc.

In our model, there are five parameters, i.e., the viscous parameter $\alpha$, BH mass $M_{\rm BH}$, dimensionless BH spin $a_*$, dimensionless accretion rate $\dot m$ [$\equiv \dot M/(M_\odot~\rm s^{-1})$], and power-law index of magnetic fields $n$. Here we concentrate on influences of $\dot m$ and $n$, so we fix the viscous parameter, BH mass, and BH spin with the typical values of $\alpha=0.1$, $M_{\rm BH} = 3~M_\odot$, and $a_*=0.9$, respectively. The BH spin $a_*$ at an extreme high value is selected because the MC effect is relatively strong in this condition \citep[e.g.][]{Lei2009,Luo2013,Song2020}. We set the accretion rates $\dot m=0.1$, 0.5, and 1, corresponding to the low, medium, and high accretion states, respectively; and set the power-law indices $n=4$ and 5\footnote{As shown in Figure 2(a) of \citet{Song2020}, in the case of $n$ = 3 with $a_*$ = 0.9, one can notice that $\theta_0$ = 0 and the outer boundary of the MC region was too small to significantly effect on the disc, so we only discussed the cases of $n$ = 4 and 5, which correspond the infinite MC regions.}, denoting the relatively incompact and compact magnetic field geometries, respectively.

\subsection{Structure}\label{structure}

\begin{figure}
\centering
\includegraphics[width=0.98\linewidth]{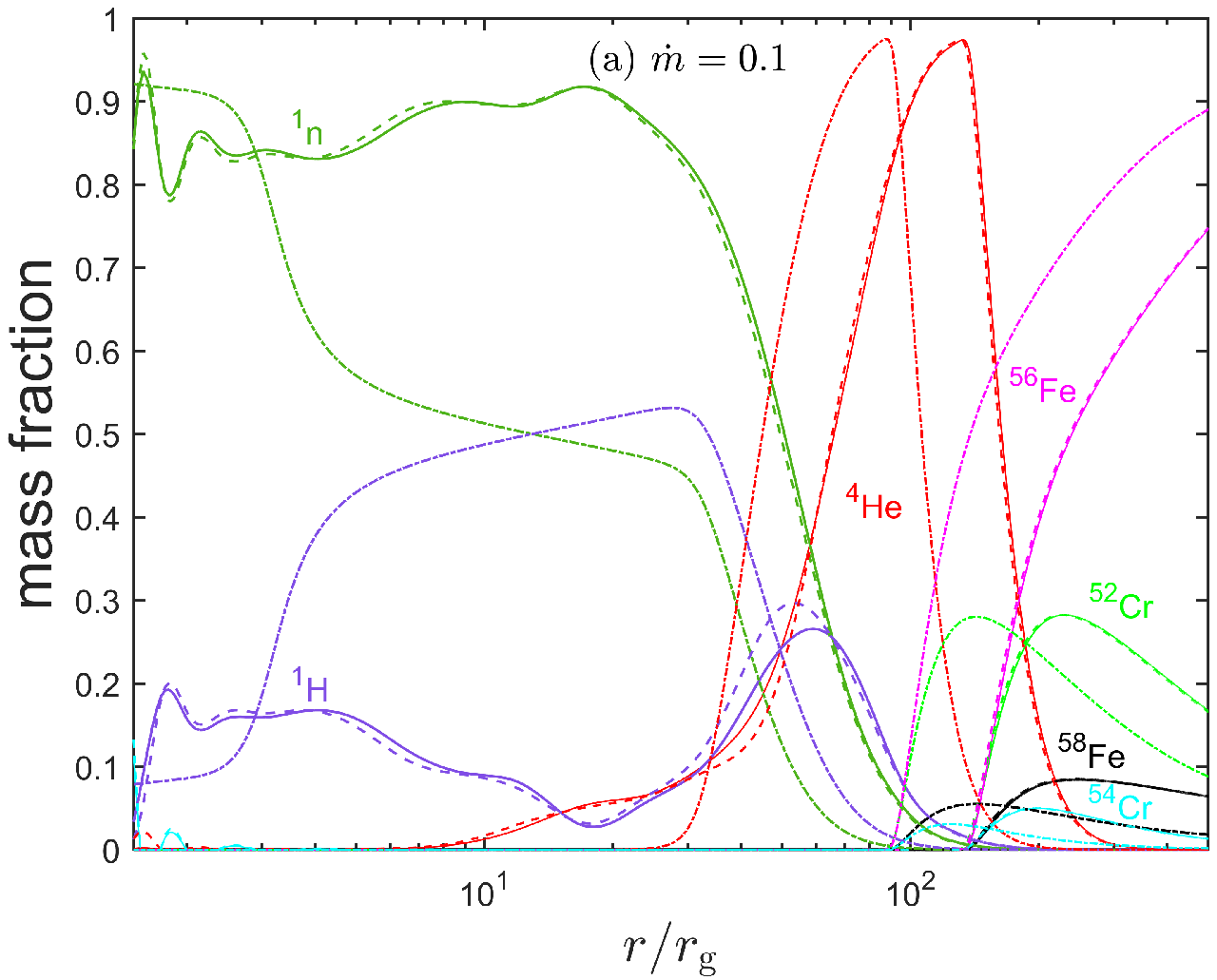}
\includegraphics[width=0.98\linewidth]{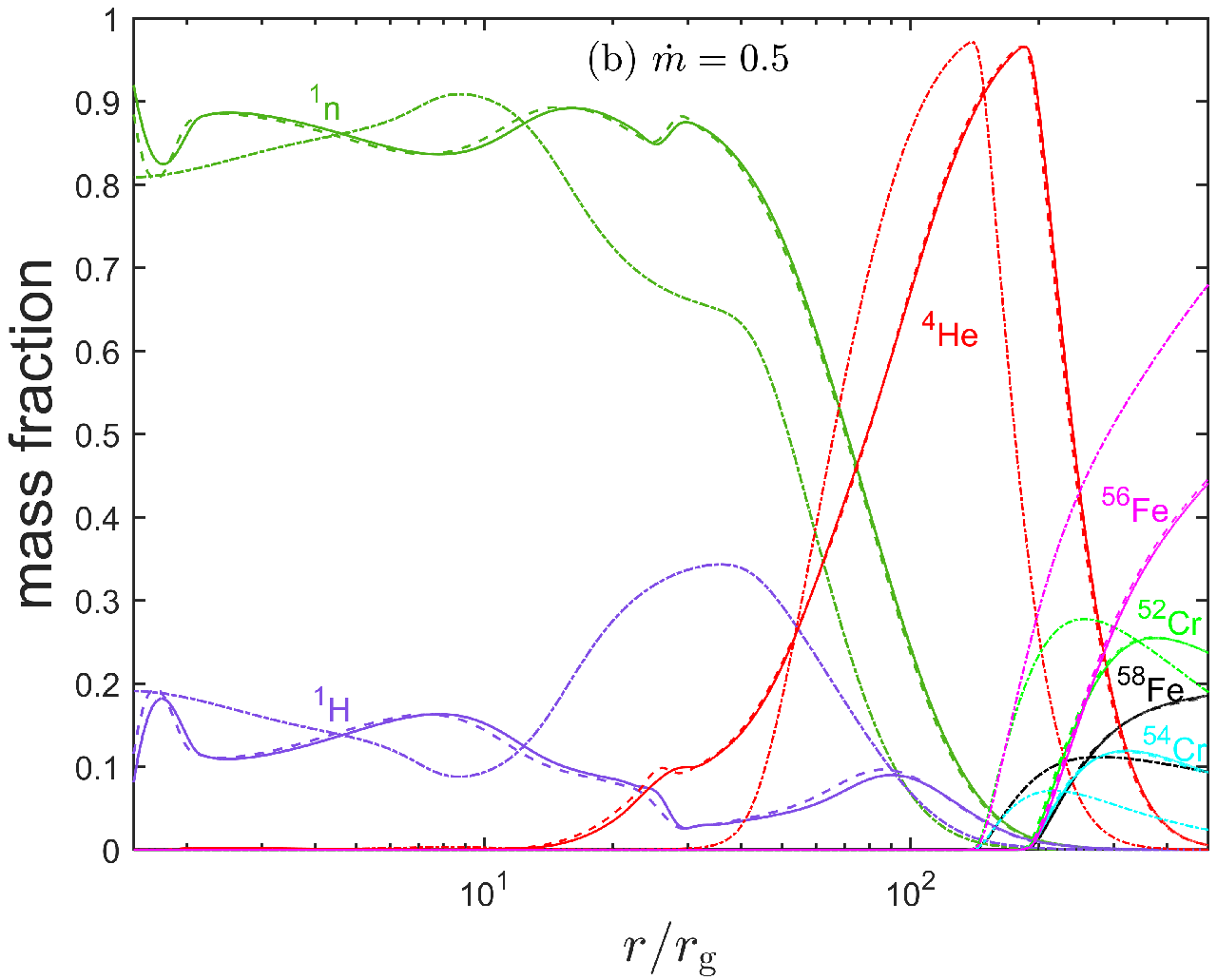}
\includegraphics[width=0.98\linewidth]{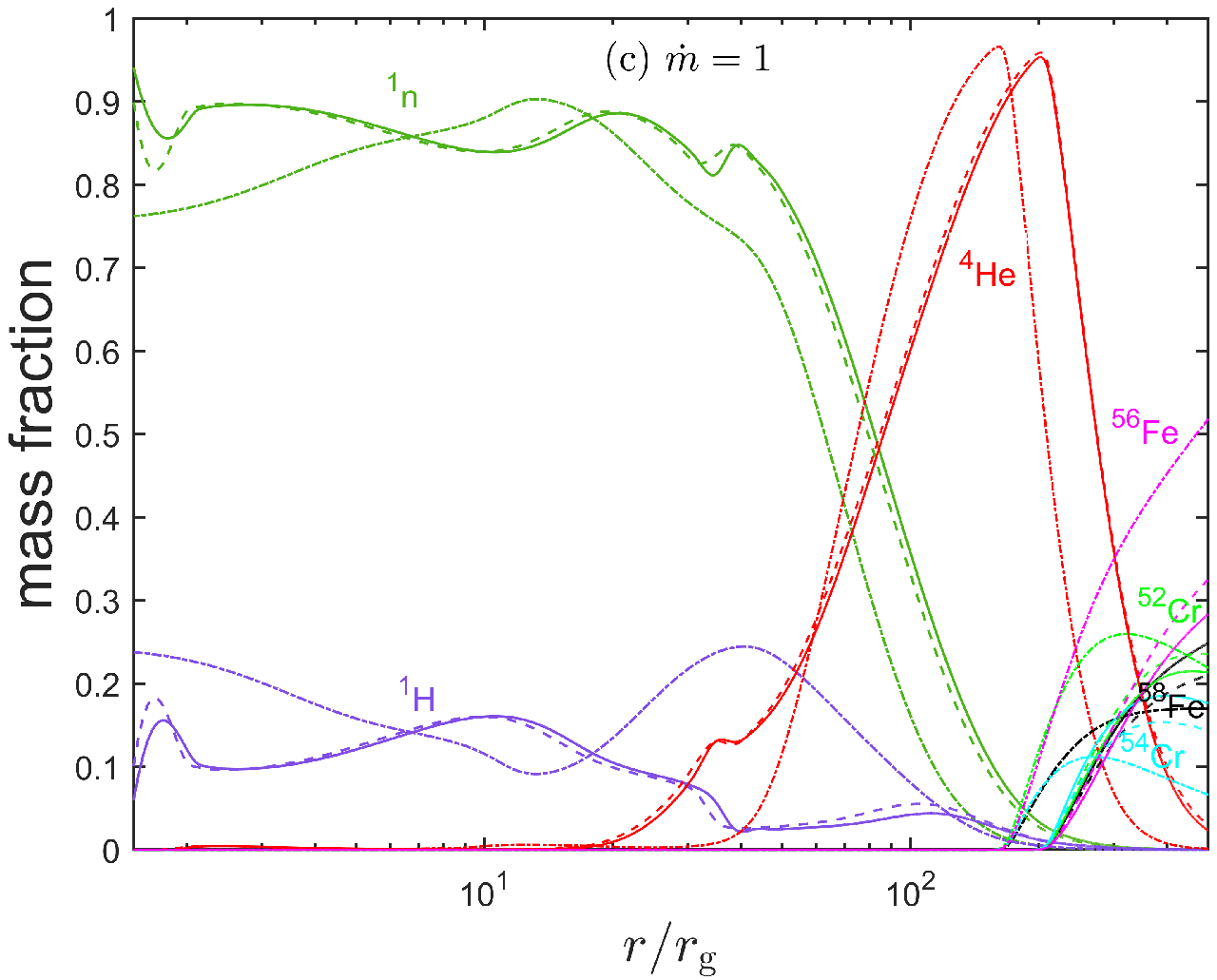}
\caption{Radial distributions of mass fractions of top seven elements, $\rm n$, $^1$H, $^4$He, $^{52}$Cr, $^{54}$Cr, $^{56}$Fe, and $^{58}$Fe for $\dot{m}=0.1$, $0.5$, and $1$, respectively. The solid and dashed lines correspond to MCNDAFs with $n=4$ and $5$ as well as the dot-dashed lines denote NDAFs.}
\label{fig_massfraction}
\end{figure}

Profiles of the density $\rho$, temperature $T$, the surface density $\Sigma$, the absolut value of the radial velocity $V_r$, the angular velocity $\Omega$, the specific angular momentum $\mathcal{L}$, and the half-thickness of disc $H$ of MCNDAFs and the counterparts of NDAFs are all shown in Figure \ref{fig_total}. The blue, red, and green lines correspond to different $\dot{m}= 0.1$, 0.5 and 1, respectively. The solid and dashed lines correspond to MCNDAFs with $n=4$ and $5$ respectively as well as the dot-dashed lines correspond to NDAFs. The density $\rho$ (or surface density $\Sigma$) and temperature $T$ of MCNDAFs are obviously higher than those of NDAFs, and reach the corresponding maximums of $\sim 10^{13}~\rm g~cm^{-3}$ and $\sim 2 \times 10^{11}~\rm K$ in the inner region of the disc when $\dot{m} = 1$. Especially, the blue solid line marginally exceeds the red solid one at $r \approx 1.5~ r_{\rm g}$ in Figure 2(c), which implies $d \dot M/ d \Sigma <0$ and indicates the viscous instability possible there. It might be an available origin of GRB variabilities, which is consistent with \citet{Lei2009}. The inflow speed $|V_r|$ of MCNDAFs are obviously slower than those of NDAFs until $r \lesssim 2~r_{\rm g}$. While there are only slight differences between MCNDAFs and NDAFs on the angular velocity $\Omega$ in the range of $2~r_{\rm g} \lesssim r \lesssim 30~ r_{\rm g}$. In the region $r/r_{\rm g}<30$, the effects of the viscosity and MC coexist and compete with each other on the transfer of angular momentum. As it is transferred from the BH to the disc by MC effects, the specific angular momentum is increased as shown in Figure \ref{fig_total}(f). Meanwhile, the temperatures of MCNDAFs at $r/r_{\rm g}<30$ are higher than those of NDAFs as shown in Figure \ref{fig_total}(b) due to the injection of additional energy through MC effects. It implies that the strength of the viscosity in the inner region of MCNDAFs is larger than that of NDAFs. Therefore, the specific angular momenta of MCNDAFs are lower than these of NDAFs. The values of the half-thickness $H$ of MCNDAFs have some fluctuations compared to these of NDAFs. All of these are caused by the additional energy and angular momentum brought by the MC process, by which the accreting gas is hindered and deposited in the inner region of the disc to result in the increasing of the temperature and density. The specific angular momentum of the deposited gas is also increased to almost the same as that in NDAFs before passing toward to BH, which can accommodate much more angular momentum with denser gas than NDAFs in the same accretion rates.

Figure \ref{fig_total1} shows the profiles of the total pressure $p$, optical depth of electron neutrinos $\tau_{\nu_e}$, electron degeneracy $\eta_e$, and electron fraction $Y_{\rm e}$ of MCNDAFs. The total pressure $p$ and the total neutrino optical depth $\tau_{\nu_{\rm e}}$ are both larger than those of NDAFs, which are consistent with the higher gas temperature and density of MCNDAFs and both result from the MC process. Especially, $\tau_{\nu_{\rm e}}$ is larger than 1 until $r \gtrsim 40~r_{\rm g}$ for $\dot{m} = 1$, which means that the inner region of MCNDAFs are totally neutrino optical thick and thermal MeV neutrinos can not escape directly from the depths of the disc in this region.

In addition, the electron degeneracy $\eta_{\rm e}$ is an important physical parameter that affects the electron fraction $Y_{\rm e}$, degeneracy pressure, and neutrino cooling \citep[e.g.][]{Liu2017b}. In Figures \ref{fig_total1}(c) and \ref{fig_total1}(d), $\eta_{\rm e}$ and $Y_{\rm e}$ of MCNDAFs are very different to those of NDAFs, which indicate the MC process has very remarkable influence on the disc microphysics. Specifically, $\eta_{\rm e}$ increases for the MCNDAF cases and its peak lies in tens of $r_{\rm g}$. On the contrary, $Y_{\rm e}$ decreases around $\sim 10~r_{\rm g}$ and tends to about 0.46 in the outer boundary. These properties will reflect in the nucleosynthesis process and related components of the disc.

\subsection{Cooling and luminosity}

Figures \ref{fig_energy1}(a-c) display the radial distributions of the neutrino cooling rates $Q_{\nu}$, the viscous heating rates $Q_{\rm{vis}}$ and the heating rates $Q_{\rm{MC}}$, respectively. The blue, red, and green lines correspond to $\dot{m}=$ 0.1, 0.5 and 1, respectively, and the solid and dashed lines correspond to MCNDAFs with $n=4$ and $5$ as well as the dot-dashed lines denote NDAFs. One can find the neutrino cooling rates $Q_{\nu}$ of MCNDAF are higher than NDAF and increase with the increasing accretion rates. The viscous rates $Q_{\rm vis}$ rise more rapidly when approach to the BH in the MCNDAF cases. The heating rates $Q_{\rm{MC}}$ in the cases of $n$ = 4 and 5 are still remarkable in the outer regions of the discs, thus the solutions of MCNDAFs and NDAFs cannot converge there as shown in Figures \ref{fig_total} and \ref{fig_total1}.

Figures \ref{fig_energy1}(d-f) show the absolute values of the advection cooling rates $\mid Q_{\rm {adv}}\mid$ in the cases of $\dot{m}= 0.1$, 0.5 and 1, the red and blue lines correspond to MCNDAFs with $n=4$ and $5$ as well as the green lines denote NDAFs, and the dash lines denote the negative values. The advection fractions change their signs to become negative due to the photodisintegration or the neutrino cooling dominated there.

In Figure \ref{fig_lvlvv}, we show the luminosities of the neutrino radiation and annihilation both for MCNDAFs and NDAFs. Obviously, there is little difference on the luminosities between the MCNDAFs with different magnetic indices but with the same accretion rates. Regardless of accretion rates, the neutrino radiation luminosities of MCNDAFs  ($\sim 10^{54}~\rm erg~s^{-1}$) are always higher about one magnitude order than those of NDAFs  ($\sim 10^{53}~\rm erg~s^{-1}$), meanwhile there is always about two magnitude order gain on the corresponding annihilation luminosities ($\sim 10^{52}-10^{54}~\rm erg~s^{-1}$ for MCNDAFs, $\sim 10^{50}-10^{52}~\rm erg~s^{-1}$ for NDAFs). This is the multiplier effect, i.e., the incident neutrino luminosities increased by 10 times, and the annihilation luminosity is proportional to the product of the incident luminosities at the colliding point, which shows the significance of MC process on the enhancement of annihilation luminosity. We also consider the neutrino trapping in our calculations because of the large increase of neutrino optical depths [see Figure \ref{fig_total1}(b)], but there is no significant effect observed in our results.

We show the spatial distribution of neutrino annihilation luminosity in Figure \ref{fig_lvvrz}. \emph{Panels} (a) and (b) plot the annihilation luminosity at different radii $r$ by integrating over the distance $z$ above the plane and at different distances above the plane $z$ by integrating the radius $r$, respectively. The line type representation is consistent with Figure \ref{fig_total}. One can note that the annihilation luminosity varies rapidly along the $z$-coordinate and the $r$-coordinate, which indicates that most of the annihilation energy escapes outward close to the inner region and along the angular momentum axis of the disc to power the relativistic jets. As mentioned before Equation (34), we adopt the Newtoniam approximation method to calculate the annihilation luminosities of NDAFs and MCNDAFs. Since the annihilation luminosities of MCNDAFs are typically higher than these of NDAFs, and neutrinos are coming from more inner parts of the disc, the relativistic effects, i.e., bending of null geodesics and gravitational redshift, should be further considered in the MCNDAF cases, which lead to the lower annihilation luminosities \citep[e.g.][]{Zalamea2011}.

\subsection{Nucleosynthesis}

In nucleosynthesis calculations, we consider more than 40 elements including free baryons, which contribute $99\%$ of accretion mass in the inner region.
Figure \ref{fig_massfraction} shows the radial distributions of mass fractions for the top seven elements, $\rm n$, $\rm ^1H$, $\rm ^4He$, $\rm ^{52}Cr$, $\rm ^{54}Cr$, $\rm ^{56}Fe$, and $\rm ^{58}Fe$ with $\dot{m}=0.1$, 0.5, and 1 respectively. For all cases, $^{56}\rm Fe$ is dominated in the outer region, $\rm ^4 He$ become dominated in the middle region, and free neutrons and protons become the final dominators in the hot and dense inner region. For different accretion rates, the ratios of free neutrons to free protons of NDAFs change significantly but this is not the case for MCNDAFs, because the MC process results in the inner region gas hotter and denser than those of NDAFs, which prevents the number of free protons from exceeding the number of free neutrons. As the same reason, the inner regions dominated by free baryons in MCNDAFs generally are larger than those in NDAFs, which results in the regions of heavy nuclei synthesis in MCNDAFs farther from BH than NDAFs.

In our conclusions, there are very rare $^{56}$Ni produced in MCNDAFs and NDAFs. According to the Urca and other adjusting processes, free protons are at a disadvantage on the numbers throughout the disc, and the synthesis condition of $^{56}$Ni is stricter than other elements of iron group, so there is hard to mass-produce $^{56}$Ni, which decay is widely considered as the origin of CCSNe \citep[e.g.][]{Liu2021}. Nevertheless, its yield might be plentiful during the adiabatic cooling process of the high-velocity disc outflows \citep[e.g.][]{Surman2006,Liu2013,Liu2021,Janiuk2014,Song2019} with a rigorous and stable condition of $Y_{\rm e} \sim 0.49-0.60$, $\rho \gtrsim 10^6~\rm g~cm^{-3}$, and $T \lesssim 5 \times 10^9~\rm K$ \citep{Seitenzahl2008}.

\section{Conclusions and discussion}\label{sec_con_dis}

In this paper, we calculate the one-dimensional global solutions of axisymmetric MCNDAFs, with the strict Kerr metric, detailed neutrino physics, different magnetic field geometries, and reasonable nucleosynthesis processes. The structures, luminosities, and elemental abundances of six cases with the different accretion rates $\dot{m}=0.1$, $0.5$, and $1$ as well as the power-law indices of magnetic fields $n=4$ and $5$ are exhibited and compared with the counterparts of NDAFs. The main results are summarized as follows.

(i) In contrast to NDAFs, the structure, thermal properties, and microphysics of MCNDAFs have been prominently impacted by the MC process especially for the inner region of the disc. There should exist the viscous instability in the inner region of MCNDAFs, which leads to the GRB variabilities.

(ii) The MC process has the significant effects on the heating mode and can increase the luminosities of neutrinos and their annihilations. The cases with the indices $n=4$ and 5 have little difference on the structure, energy equilibrium, luminosity, and components.

(iii) The mass fractions of $^{56}$Fe, $^4$He, and free neutrons are dominated in turn in the outer, middle, and inner regions of MCNDAFs. The MC process will drive off the heavy nuclei synthesis to the outside of MCNDAFs farther than that of NDAFs. Almost no $^{56}$Ni appears in the outer disc.

Therefore, the MC effect plays an important role to rebuild the structure, luminosity, and components of MCNDAFs. Moreover, although the NDAF is one of the popular central engine model of GRBs, only a small fraction of neutrinos can annihilate in the outside space of the disc to power GRBs and other following radiation processes, so there still leaves over some issues related to the much higher energy requirements from some observed luminous GRBs. The MC process is one of the priority options to dramatically enhance the neutrino annihilation luminosity, and might attach some other advantageous effects, such as arising the possible instability in the inner disc \citep[e.g.][]{Janiuk2007,Lei2009}, redistributing the angular momentum of the disc \citep[e.g.][]{Janiuk2008,Liu2012a}, and launching the episodic Poynting fluxes via magnetic reconnections \citep[e.g.][]{Yuan2012,Liu2017b}.

Since the drastic hyperaccretion process appears in the BH-NDAF systems, the evolutions of the central BH mass and spin should be further considered \citep[e.g.][]{Janiuk2008,Janiuk2010,Song2015,Liu2021}. Of course, the strong disc outflows and MC process might greatly relieve this situation. Moreover, the electron fraction $Y_{\rm e}$ of MCNDAFs tends to the relative proton-rich phase as well as in NDAF cases \citep[e.g.][]{Xue2013}, which is naturally applicable to describe the initial fallback hyperaccretion in the massive collapsar scenario. As mentioned above, the $^{56}$Ni yield in the disc is very scarcity also caused by the self-consistent $Y_{\rm e}$ condition and the following ``missing'' protons via Urca process; but the disc outflows will change the structure and components of the disc and rewrite the equations related to $Y_{\rm e}$, then could be the rich $^{56}$Ni mines as well as their host CCSNe, which will prominently contribute on the chemical evolution in the Universe \citep[e.g.][]{Liu2021}. In future work, a compulsive boundary condition should be included to investigate the properties of the time-dependent NDAFs with magnetic fields and outflows lying in the various circumstances with different element abundances.

\section*{Acknowledgements}

We thank the anonymous referee for helpful suggestions. This work was supported by the National Natural Science Foundation of China under grants 12173031, 11822304, and 11373002, and the science research grants from the China Manned Space Project with No. CMS-CSST-2021-B11.

\section*{Data availability}

The data underlying this article will be shared on reasonable request to the corresponding author.

\end{document}